\documentclass[reprint,aps,pra,superscriptaddress,english,nolongbibliography]{revtex4-2}

\usepackage[utf8]{inputenc}
\usepackage{natbib}

\usepackage[dvipsnames]{xcolor}
\usepackage{graphicx}
\usepackage[colorlinks=true,citecolor=blue,urlcolor=blue,linkcolor=blue]{hyperref}
\usepackage{braket}
\usepackage{amssymb,amsmath,amsfonts,mathrsfs,fancyhdr}

\usepackage[caption=false,subrefformat=parens,labelformat=empty]{subfig}

\DeclareMathOperator{\typ}{typ}

\begin{document}


\title{Spatio-temporal heterogeneity of entanglement in many-body localized systems}
\author{Claudia Artiaco}
\email{artiaco@kth.se}
\affiliation{Scuola Internazionale Superiore di Studi Avanzati (SISSA), via Bonomea 265, 34136, Trieste, Italy}
\affiliation{The Abdus Salam International Center for Theoretical Physics, Strada Costiera 11, 34151, Trieste, Italy}
\affiliation{INFN Sezione di Trieste, Via Valerio 2, 34127 Trieste, Italy}
\affiliation{Department of Physics, KTH Royal Institute of Technology, Stockholm 106 91, Sweden}
\author{Federico Balducci}
\affiliation{Scuola Internazionale Superiore di Studi Avanzati (SISSA), via Bonomea 265, 34136, Trieste, Italy}
\affiliation{The Abdus Salam International Center for Theoretical Physics, Strada Costiera 11, 34151, Trieste, Italy}
\affiliation{INFN Sezione di Trieste, Via Valerio 2, 34127 Trieste, Italy}
\author{Markus Heyl}
\affiliation{Max Planck Institute for the Physics of Complex Systems, N\"othnitzer Stra{\ss}e 38, D-01187, Dresden, Germany}
\affiliation{Theoretical Physics III, Center for Electronic Correlations and Magnetism, Institute of Physics, University of Augsburg, D-86135 Augsburg, Germany}
\author{Angelo Russomanno}
\affiliation{Max Planck Institute for the Physics of Complex Systems, N\"othnitzer Stra{\ss}e 38, D-01187, Dresden, Germany}
\affiliation{Scuola Superiore Meridionale, Università di Napoli Federico II, Largo San Marcellino 10, I-80138 Napoli, Italy}
\author{Antonello Scardicchio}
\affiliation{The Abdus Salam International Center for Theoretical Physics, Strada Costiera 11, 34151, Trieste, Italy}
\affiliation{INFN Sezione di Trieste, Via Valerio 2, 34127 Trieste, Italy}


\begin{abstract}
    We propose a spatio-temporal characterization of the entanglement dynamics in many-body localized (MBL) systems, which exhibits a striking resemblance with dynamical heterogeneity in classical glasses. Specifically, we find that the relaxation times of local entanglement, as measured by the concurrence, are spatially correlated yielding a dynamical length scale for quantum entanglement. As a consequence of this spatio-temporal analysis, we observe that the considered MBL system is made up of dynamically correlated clusters with a size set by this entanglement length scale. The system decomposes into compartments of different activity such as active regions with fast quantum entanglement dynamics and inactive regions where the dynamics is slow.
    We further find that the relaxation times of the on-site concurrence  become broader distributed and more spatially correlated, as disorder increases or the energy of the initial state decreases. Through this spatio-temporal characterization of entanglement, our work unravels a previously unrecognized connection between the behavior of classical glasses and the genuine quantum dynamics of MBL systems.
\end{abstract}

\maketitle

\section{Introduction}

The assumption of local equilibrium is at the core of statistical mechanics: even if isolated from the rest of the universe, a generic many-body system is expected to act as a thermal bath for itself, quickly driving the statistics of local observables to the Gibbs ensemble, by means of classical~\cite{Vulpiani,lichtenberg2013regular,berry1978regular} or quantum~\cite{Polkovnikov2011Colloquium,d2016quantum} chaos. The situations in which ergodization fails, and the system persists in non-thermal states for all relevant times, are therefore of paramount interest, both at the classical and the quantum level \cite{deutsch1991quantum,srednicki1994chaos,d2016quantum}.

Glasses are a prototypical example of classical systems that remain trapped in metastable states for all experimentally accessible time scales \cite{angell2000relaxation,debenedetti2001supercooled,cavagna2009supercooled,berthier2011theoretical,biroli2013perspective}. In the quantum realm, after the seminal work \cite{basko2006metal}, it has become clear that isolated, disordered many-body systems can elude thermal equilibrium even at \emph{infinite} time. The existence of this nonergodic phase, coined \emph{many-body localized} (MBL), was found analytically \cite{gornyi2005interacting,basko2006metal,imbrie2016many,imbrie2016diagonalization} and numerically in a vast set of microscopic models \cite{oganesyan2007localization,znidaric2008many,pal2010many,berkelbach2010conductivity,deLuca2013,kjall2014many,luitz2015many,brenes2018many}, and observed in ultracold-atom experiments \cite{schreiber2015observation,bordia2016coupling,smith2016many}. 
The lack of ergodicity in the MBL phase has been linked to the existence of an extensive number of local integrals of motion (LIOMs) \cite{serbyn2013local,huse2014phenomenology,ros2015integrals,imbrie2016review,chandran2016many,varma2019length,peng2019comparing}, by which one can construct a phenomenological model known as the \emph{$l$-bit model}. The $l$-bit model qualitatively captures the features of the MBL phase, such as slow decay of correlation functions, area-law entanglement for eigenstates, and slow spreading of entanglement after a quantum quench~\cite{bardarson2012unbounded,serbyn2013universal,Swingle2013Simple,serbyn2014quantum,Serbyn2016PowerLaw,Znidaric18}.

While entanglement can be completely characterized for two qubits~\cite{nielsen_chuang_2010,Wootters98,hill1997entanglement}, this becomes more challenging in the many-body context incorporating a multitude of facets~\cite{facchi2008maximally,facchi2010multipartite,Amico08,laflorenice2016}. The numerous studies on the entanglement growth in the MBL phase~\cite{pietracaprina2017entanglement,montangero2006entanglement,bardarson2012unbounded,znidaric2008many,serbyn2013universal} are mainly focused on global properties, employing measures such as the entanglement entropy, purity, quantum Fisher, or mutual information~\cite{smith2016many,de2017quantum, de2019efficiently,herviou2019multiscale}, or total correlations~\cite{nielsen_chuang_2010,goold2015total}. 

In this work, we go beyond those approaches, focusing on the local properties of entanglement, therefore aiming at characterizing its spreading in a more detailed way. We focus on the combined temporal and spatial behavior of the local entanglement as measured by the concurrence, which has been used to characterize entanglement in MBL before (but only at the global level) \cite{iemini2016signatures}, and which can be measured experimentally~\cite{jurcevic2014quasiparticle,fukuhara2015spatially}. Our starting point is the finding that the dynamics of the concurrence among couples of spins or $l$-bits is highly heterogeneous, with a wide range of different relaxation times. To quantify this observation, we investigate the distribution of the relaxation times $\tau_i$ of the concurrence in the $l$-bit model, and describe its properties in a wide range of parameters and initial-state energies. Our main findings are the following. First, the strong fluctuations of entanglement at the local level manifest as a power-law-tailed probability distribution for $\tau_i$ which, in turn, is at the origin of the known power-law decay of the \emph{average} concurrence~\cite{iemini2016signatures}. Second, we show that the width of such distribution increases as disorder increases or energy decreases. Third, we illustrate that the local relaxation times $\tau_i$ are spatially correlated, with the correlations growing as disorder increases or energy decreases: this is a counter-intuitive result, since, naively, one might expect the correlations to increase when approaching the delocalization transition by \emph{lowering} the disorder. It is worth stressing that this latter result suggests that the associated entanglement correlation length represents a previously unrecognized length scale in the MBL phase.

These results provide a connection with the so-called \emph{dynamical heterogeneity} observed in classical amorphous materials and spin glasses~\cite{glotzer1998dynamical,ediger2000spatially,glotzer2000spatially,cavagna2009supercooled,berthier2011theoretical,berthier2011dynamical,berthier2011dynamic}, adding one more point of contact between the phenomenology of glasses and that of MBL systems~\cite{Schiulaz2014Ideal,abanin2019colloquium,artiaco2021quantum,artiaco2021signatures}. One speaks about dynamical heterogeneity when each local degree of freedom presents an autocorrelation function that decays in time with a different functional form, thus strong spatio-temporal fluctuations are present in the system. It has already been argued that dynamical heterogeneity in classical glass models has a quantum counterpart and, at ultra-low temperatures, can be induced by quantum fluctuations~\cite{olmos2012facilitated,nussinov2013mapping,lan2018quantum,Gopalakrishnan20}. However, previous studies have investigated quantum glass systems modeled on a classical counterpart. Here, instead, we adopt a different approach: we borrow the theoretical tools of classical glass theory in order to explore the purely quantum MBL 
phase and study the features of entanglement, which is a genuine quantum feature with no classical analog. In this perspective, the observed similarities in the heterogeneous behavior of local correlations, which are quantum in the MBL case and classical for glasses, appear unexpected.

Importantly, entanglement heterogeneity persists also outside the MBL phase. Here, we choose to focus on the deep MBL phase only because, thanks to the $l$-bit model, our numerics can reach larger system sizes, and we can also get some analytical insight. We defer the detailed study of entanglement heterogeneity out of the MBL phase, and across the MBL-thermal transition, to future studies.

The paper is organized as follows. In Sec.~\ref{sec:model}, we describe the models under study: the XXZ chain with disorder and the $l$-bit model deep in the MBL phase. In Sec.~\ref{sec:methods}, we introduce the quantities we investigate, i.e.\ the on-site concurrences, and explain how we evaluate their relaxation times and the correlations among such times. In Sec.~\ref{sec:distributions}, we show the results on the distribution of the relaxation times: in the MBL phase, we find that they have a power-law tail and that we can analytically predict the dependence of the typical relaxation-time value on the disorder strength. In Sec.~\ref{sec:correlations}, we discuss the results for the correlation function of the relaxation times, showing that their correlation length increases with the disorder strength. Finally, in Sec.~\ref{sec:conclusions} we draw our conclusions and discuss future research perspectives.

\section{Model}
\label{sec:model}

We aim at studying the general properties of the spatio-temporal entanglement dynamics of MBL systems. For this purpose, we focus on an effective description in terms of LIOMs, which allows us to access the nonequilibrium real-time dynamics of MBL systems for long times and large system sizes. Deep in the MBL phase, Hamiltonians of short-range interacting quantum spin-1/2 degrees of freedom can be diagonalized through a quasi-local unitary transformation~\cite{imbrie2016many,imbrie2016review}, yielding a representation of the model in so-called $l$-bit form: 
\begin{equation}
    H_{l\text{-bit}} = \sum_{i=1}^L h_i \sigma_i^z + \sum_{i,j=1}^L J_{ij} \sigma_i^z \sigma_j^z + \dots
    \label{eq:lbit_model}
\end{equation}
where $\{\sigma^x_i,\sigma^y_i,\sigma^z_i\}$ are the localized spin-1/2 operators associated with the LIOMs. We neglect further terms in the Hamiltonian which comprise $n$-body interactions with $n \geq 3$, which is a controlled approximation for weakly interacting spins in the original microscopic model. From analytical results \cite{serbyn2013local,imbrie2016many,ros2015integrals,imbrie2016review}, it is known that the interactions $J_{ij}$ are exponentially suppressed with the distance $r_{ij}$ between localization centers. To achieve a model-independent effective description, we parameterize the $l$-bit model as follows. We assume that the $h_i$ are independent identically distributed random fields, with a uniform distribution over $[-h,h]$, and that the $J_{ij}$ are uncorrelated Gaussian variables of zero average and standard deviation $J_0 e^{-r_{ij}/\kappa}$. For numerical purposes, we set $h=J_0=1$.

The particular advantage of the $l$-bit model~\eqref{eq:lbit_model} is that it allows us to perform analytical estimates of few-body observables, and to efficiently compute them numerically, reaching system sizes up to $L=140$ spins for very long times. To give an example, Eq.~\eqref{eq:app_xx_correlation} in App.~\ref{sec:app_typical_estimate} shows how to compute a two-site correlation function in $O(N)$ steps. We refer the reader to Refs.~\cite{serbyn2013universal,Swingle2013Simple,serbyn2014quantum,iemini2016signatures,Znidaric18,artiaco2021signatures} for more general observables.

It is worth stressing that the $l$-bits become closer and closer to the physical spins as the disorder increases, ultimately coinciding asymptotically at infinite disorder \cite{varma2019length,peng2019comparing}. Thus, at small values of $\kappa$ (i.e.\ large disorder strength), one can safely consider the $l$-bits as uniformly spaced on a chain, and compute the distances between them as $r_{ij} = |i-j|$, $i,j=1,2,\dots,L$. Our numerical results are obtained exactly in this strongly localized regime, deep in the MBL phase. It is important to note that the effective model allows us to tune: i) the interaction decay length $\kappa$ (equivalent to varying the disorder strength); and ii), the initial condition, i.e.\ the energy density at which we probe the system's properties. Concerning the latter parameter, we choose as initial state of the dynamics a product state in the \emph{effective} spin basis: 
\begin{equation}
    | \psi_0 \rangle = \bigotimes_{i=1}^{L} \big(A_i | \! \Uparrow \rangle_i + B_i |\! \Downarrow \rangle_i \big),
    \label{eq:lbit_initial_state}
\end{equation}
where $|\! \Uparrow \rangle_i$,$|\!\Downarrow \rangle_i$ are the eigenstates of $\sigma_i^z$, and $|A_i|^2 + |B_i|^2 = 1$. Employing Eq.~\eqref{eq:lbit_initial_state}, the system is initially prepared in a superposition of eigenstates. Moreover, Eq.~\eqref{eq:lbit_initial_state} provides us with the flexibility to tune the coefficients $A_i$ and $B_i$ such that we can vary the initial-state energy expectation value $E := \langle \psi_0 | H_{l\text{-bit}} | \psi_0 \rangle$. This tuning can be achieved using a classical simulated annealing algorithm (see Appendix~\ref{sec:app_initial_E} for details), and allows us to explore different regions of the energy spectrum. We measure $E$ in units of the standard deviations of $h_i$ and $\sum_j J_{ij}$, defining the dimensionless energy density $\varepsilon := (E/N)/\sqrt{h^2/3 + 2J_0^2/(e^{2/\kappa}-1)}$. Notice that $\varepsilon=0$ corresponds to the center of the spectrum, while $\varepsilon \approx -1$ to the ground state (more details in App.~\ref{sec:app_initial_E}). Let us remark that the localization properties of MBL systems depend on the energy of the considered state and are stronger near the edges of the spectrum \cite{luitz2015many,alet2018many}.

For small system sizes, we will compare the results of the effective model with a full microscopic calculation for the spin-1/2 XXZ chain with random fields:
\begin{equation}
    H_{\mathrm{XXZ}} = \sum_{i=1}^{L-1} \left[\frac{J}{2}(S^+_iS^-_{i+1} + \mathrm{h.c.}) + VS^z_iS^z_{i+1}\right] + \sum_{i=1}^L \Delta_i S^z_i,
    \label{eq:XXZ_model}
\end{equation}
where $J=V=1$ (unless otherwise stated) and $\Delta_i$ are random variables uniformly distributed over $[-\frac{W}{2},\frac{W}{2}]$. For $W_c \simeq 7 \pm 2$, this model exhibits an MBL transition~\cite{pal2010many} (see Refs.~\cite{varma2019length,peng2019comparing} for the relation between $W$ and the effective model parameters $h$, $\kappa$ and $J_0$). When employing~\eqref{eq:XXZ_model}, we probe the centre of the energy spectrum initializing the system in a N\'{e}el state $|\psi_0\rangle = |\uparrow \downarrow \uparrow \downarrow \dots \rangle$ where $\uparrow, \downarrow$ indicates the physical spin basis, and average the results over different disorder realizations. Notice that the XXZ model can be employed to explore the presence of entanglement heterogeneity also in the thermal phase; however, due to the smallness of the accessible system sizes, we will not investigate thoroughly the thermal phase in the present work. We defer such investigation to future studies.

\section{Methods}
\label{sec:methods}

For the purpose of exploring the spatio-temporal heterogeneity of entanglement in MBL systems, we concentrate on the two-site concurrence, which quantifies the pairwise entanglement between two qubits \cite{hill1997entanglement,Wootters98,Amico08,iemini2016signatures}. For two spins-1/2 located at lattice sites $i$ and $j$, the concurrence is defined as \cite{Wootters98,Amico08}
\begin{equation}
\label{eq:general_conc}
    C_{i,j} := \max{ \{0, \, \lambda_1 - \lambda_2 - \lambda_3 - \lambda_4 \}}.
\end{equation}
$\lambda_a^2$ are the eigenvalues of the matrix $R_{ij} = \rho_{ij} (\sigma_y \otimes \sigma_y) \rho^*_{ij} (\sigma_y \otimes \sigma_y)$ sorted in descending order, where $\rho_{ij}$ is the two-site reduced density matrix, and the complex conjugation is done in the standard computational basis.

While the general formula~\eqref{eq:general_conc} must be applied to the $l$-bit model, for the microscopic XXZ Hamiltonian the concurrence can be computed more easily. Since the dynamics conserves the $z$ component of the total magnetic field, $S^z_{\text{tot}}$, and we initialize the system in the N\'eel state, having $S^z_{\text{tot}}=0$, Eq.~\eqref{eq:general_conc} can be simplified as \cite{yu2007evolution}
\begin{equation}
\label{eq:concurrence_XXZ}
    C_{i,j} = 2\max\left[0, \left| \braket{S_i^+ S_j^-} - \sqrt{P_{++} P_{--}} \right| \right]
\end{equation}
with
\begin{equation}
    P_{\pm\pm} \equiv \frac{1}{4} \pm \frac{1}{2} \left( \braket{S_i^z} + \braket{S_j^z} \right) + \braket{S_i^z S_j^z}.
\end{equation}
and $\langle \bullet \rangle = \langle \psi(t) | \bullet | \psi(t) \rangle$, where $\psi(t)$ is the state of the system at time $t$.

It has been shown that in MBL systems the concurrence averaged over all couples $i,j$ decays in time as a power law \cite{iemini2016signatures}, while it decays exponentially fast in ergodic systems. Our key goal is to establish a more detailed spatio-temporal analysis of the concurrence beyond its averaged value. Specifically, we wish to investigate, in MBL systems, the relationship between the late-time, power-law behavior of the average concurrence and the relaxation of the concurrence at the local level. Because of the presence of quenched disorder, we expect to observe that the relaxation of the two-site entanglement is \emph{heterogeneous}, i.e.\ it has a different functional form and characteristic time scale in distinct spatial regions. We aim at verifying and quantifying such entanglement heterogeneity. Similar questions have been already explored in the framework of classical glasses; in this work, we will adapt some tools and ideas developed in that context to the case of quantum localized systems.

In this perspective, we define a local \emph{on-site} concurrence as
\begin{equation}
    \label{eq:one-site_concurrence}
    C_i(t) :=  \sum_j C_{i,j}(t) \, ,
\end{equation}
quantifying the total amount of two-qubit entanglement of $i$ with all the other lattice sites. In the case of the XXZ model, we find that $C_{i,j} \simeq 0$ for $|i-j|>1$, so that we can trade the sum in Eq.~\eqref{eq:one-site_concurrence} with the nearest-neighbour term: $C_i(t) := C_{i,i+1}(t)$ (see also~\cite{iemini2016signatures}).

\begin{figure}
    \centering
    \includegraphics[width=\columnwidth]{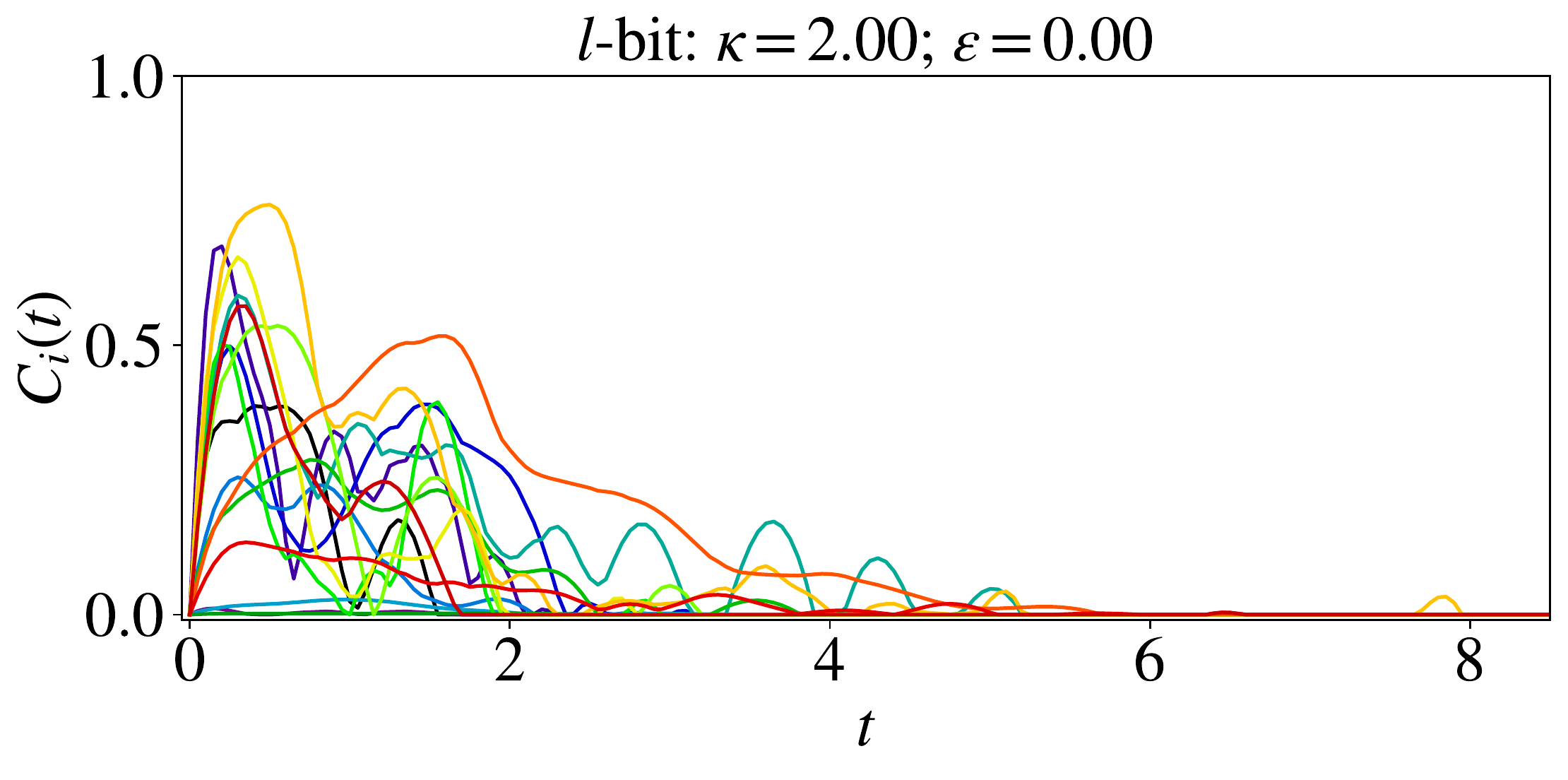}
    \caption{Single instance for the on-site concurrences $C_i(t)$, defined in Eq.~\eqref{eq:one-site_concurrence}. Heterogeneity in the entanglement spreading can be clearly seen: the curves $C_i(t)$ are described by different functional forms, and decay to zero on different time scales. The chain length is of $L=80$ sites. The concurrence of one site every three is plotted to enhance readability.}
    \label{fig:allSites}
\end{figure}

For large systems and for a single disorder realization, we find that the local concurrence defined in Eq.~\eqref{eq:one-site_concurrence} typically decays to zero on a certain time scale, and then definitely remains so (see Fig.~\ref{fig:allSites}). This motivates us to define the local relaxation time as 
\begin{equation}
    \label{eq:def_tau}
    \tau_i := t_0 \, e^{\langle \ln (t/t_0) \rangle_C} := t_0 \exp \frac{\int_0^{t_{\rm fin}}\ln(t/t_0) \, C_i(t) dt}{\int_0^{t_{\rm fin}}C_i(t) dt},
\end{equation}
where $t_0=J_0^{-1}$. Notice that $C_i \geq 0$, so the averages above are well-defined and independent of $t_0$ (in the thermodynamic limit). The definition~\eqref{eq:def_tau} employs the logarithm $\ln(t/t_0)$; this ensures that $\tau_i$ is a good estimator of the \emph{typical} time scale of the relaxation time of the concurrence even if $C_i(t)$ decays very slowly \footnote{We verified that, upon changing the definition of $\tau_i$, e.g. with $\tau_i := \langle t \rangle_C = \int_0^{t_{\rm fin}} t \, C_i(t)  dt / \int_0^{t_{\rm fin}}C_i(t) dt$ or $\tau_i := \max \{ t| C_i(t)>0 \}$, our findings do not qualitatively change.}. Notice that typically, for \emph{finite} systems, $C_i(\infty) \simeq O(2^{-L})$ \cite{artiaco2021signatures}: thus, the function $C_i(t)$ might be interpreted as a probability distribution over $\mathbb{R}$ only in the thermodynamic limit.

Our aim is to estimate the distribution function of $\tau_i$ exactly in the thermodynamic limit. Such limit can be approached by fixing the maximum simulation time of the dynamics $t_{\rm fin}$ and increasing $L$, until convergence is reached. We find from our numerics that this typically happens for $L\gtrsim 30$, which is achievable in the $l$-bit but not for microscopic Hamiltonians by means of exact diagonalization. Indeed, for the microscopic XXZ Hamiltonian in the MBL phase and for finite $L$, one finds a spurious peak in the distribution of $\tau_i$ due to those realizations of $C_i(t)$ which are still nonzero at the final evolution time $t_{\text{fin}}$. This is precisely due to the fact that, for small $L$ and whatever the choice of $t_{\text{fin}}$, there will be always a finite number of such nonvanishing realizations. Unfortunately, for the XXZ model, we cannot consider system sizes larger than $L=20$. This represents a crucial argument for the use of the effective $l$-bit model.

Let us emphasize again that the average concurrence in the MBL phase decays as a power law, while the on-site concurrence vanishes after a finite time, as depicted in Fig.~\ref{fig:allSites}. Thus, the power-law decay of the average is nothing but a consequence of the heterogeneous behavior at the local level. This can be also captured by a simple analytical model, as sketched in the following. Let us schematize the on-site concurrence as a step function $f_\tau(t)=\theta(\tau-t)$, where $\tau$ is a random variable drawn from a power-law-tailed distribution. Assuming $P(\tau) = \mathcal{N} \theta(\tau-\tau_0) \tau^{-\gamma}$, with $\gamma>1$ and $\mathcal{N}$ a normalization constant, one gets a disorder average $\braket{f(t)} = (\tau_0 /t )^{\gamma-1}$ that decays as a power law. This shows that, when $f_\tau(t)$ has a simple form and vanishes after a finite time in each disorder realization, and when the relaxation times have a long-tailed power-law distribution, the average $\braket{f(t)}$ decays as a power law. We expect this argument to be at the origin of the power-law decay of the average concurrence in MBL systems.

While the $\tau_i$'s provide us with temporal information of the entanglement dynamics, we are further interested in the spatial component. For that purpose we quantify the spatial correlations of the local relaxation time via (see also Ref.~\cite{glotzer1998dynamical})
\begin{equation}
    \label{eq:def-Gtau}
    G_{\tau}(r) := \overline{\left[ \frac{\langle \tau_i \tau_j \rangle_{\textrm{is}} - \langle \tau_i \rangle_{\textrm{is}} \langle \tau_j \rangle_{\textrm{is}}}{\langle \tau_i^2 \rangle_{\textrm{is}} - \langle \tau_i \rangle_{\textrm{is}}^2} \right]_{|i-j|=r}} \, ,
\end{equation}
where $\langle \bullet \rangle_{\textrm{is}}$ denotes the average over different initial states, $[\bullet]_{|i-j|=r}$ the average over all sites $i,j$ separated by a distance~$r$, and $\overline{\bullet}$ the average over different disorder realizations \footnote{Notice that the averages have to be taken in the proper order: first $\langle \bullet \rangle_{\textrm{is}}$, second $[\bullet]_{|i-j|=r}$, finally $\overline{\bullet}$.}. In Appendix~\ref{self:app}, we show that $G_{\tau}(r)$ as defined in Eq.~\eqref{eq:def-Gtau} is very robust to finite-size effects and disorder fluctuations: it is a self-averaging quantity. From our numerical simulations, we find that $G_{\tau}(r)$ experiences in general a stretched-exponential decay as a function of $r$. This allows us to define a length scale $\eta_\tau$ by performing a fit of the form $\log G_{\tau}(r) \sim a + (r/\eta_\tau)^b$ for some suitable $a$ and $b$. The length $\eta_\tau$ quantifies the distance over which the local entanglement relaxation is spatially correlated, i.e.\ it gives the size of the typical clusters of fast or slow entangling spins. In App.~\ref{self:app}, we show also that $\eta_\tau$ is almost independent of the system size for $L \geq 40$.

\begin{figure}[t]
    \centering
    \subfloat[]{\includegraphics[width=\columnwidth]{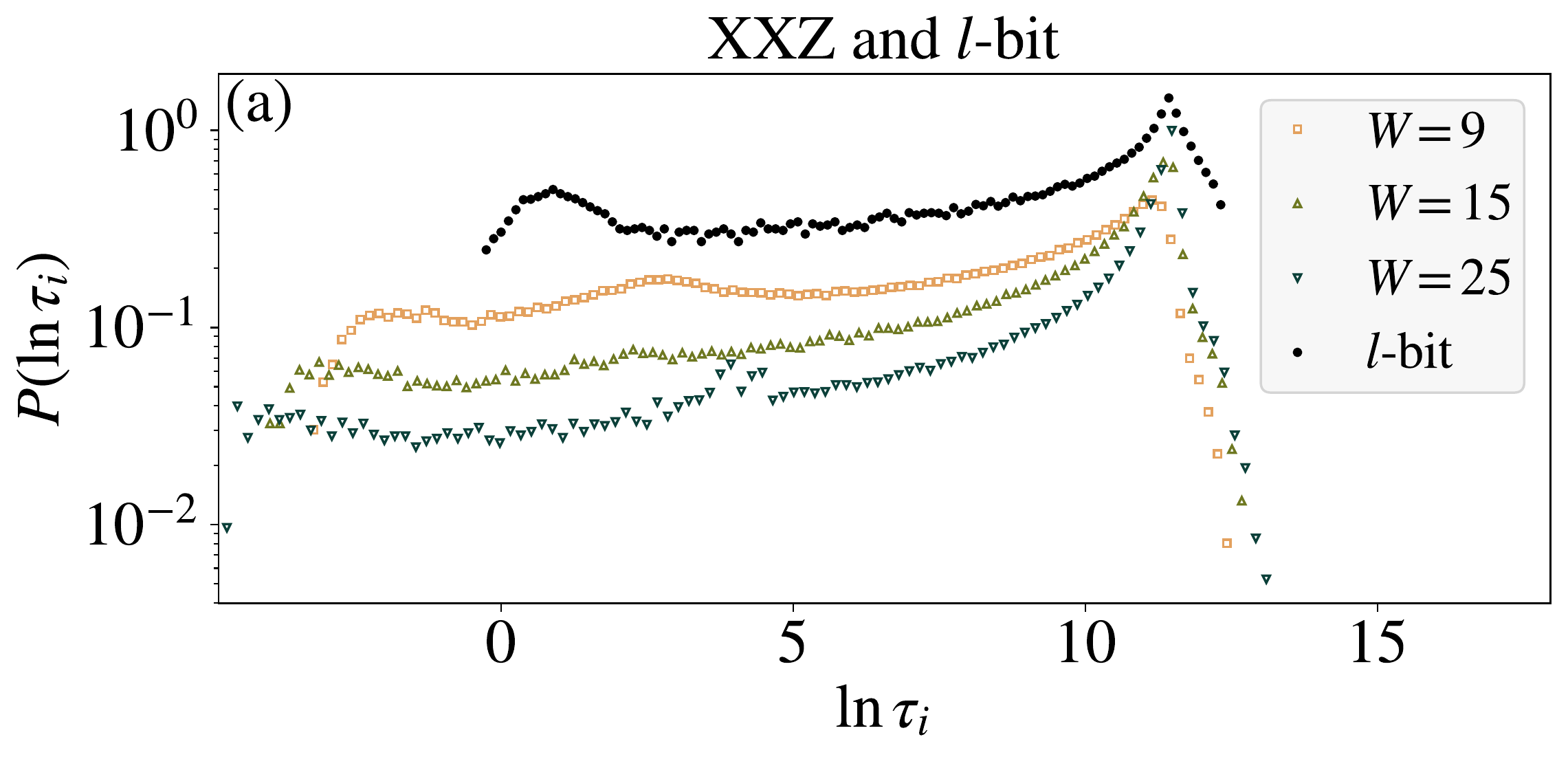}
    \label{fig:histo-tau-XXZ}}
    \vspace{-8mm}
    \subfloat[]{\includegraphics[width=\columnwidth]{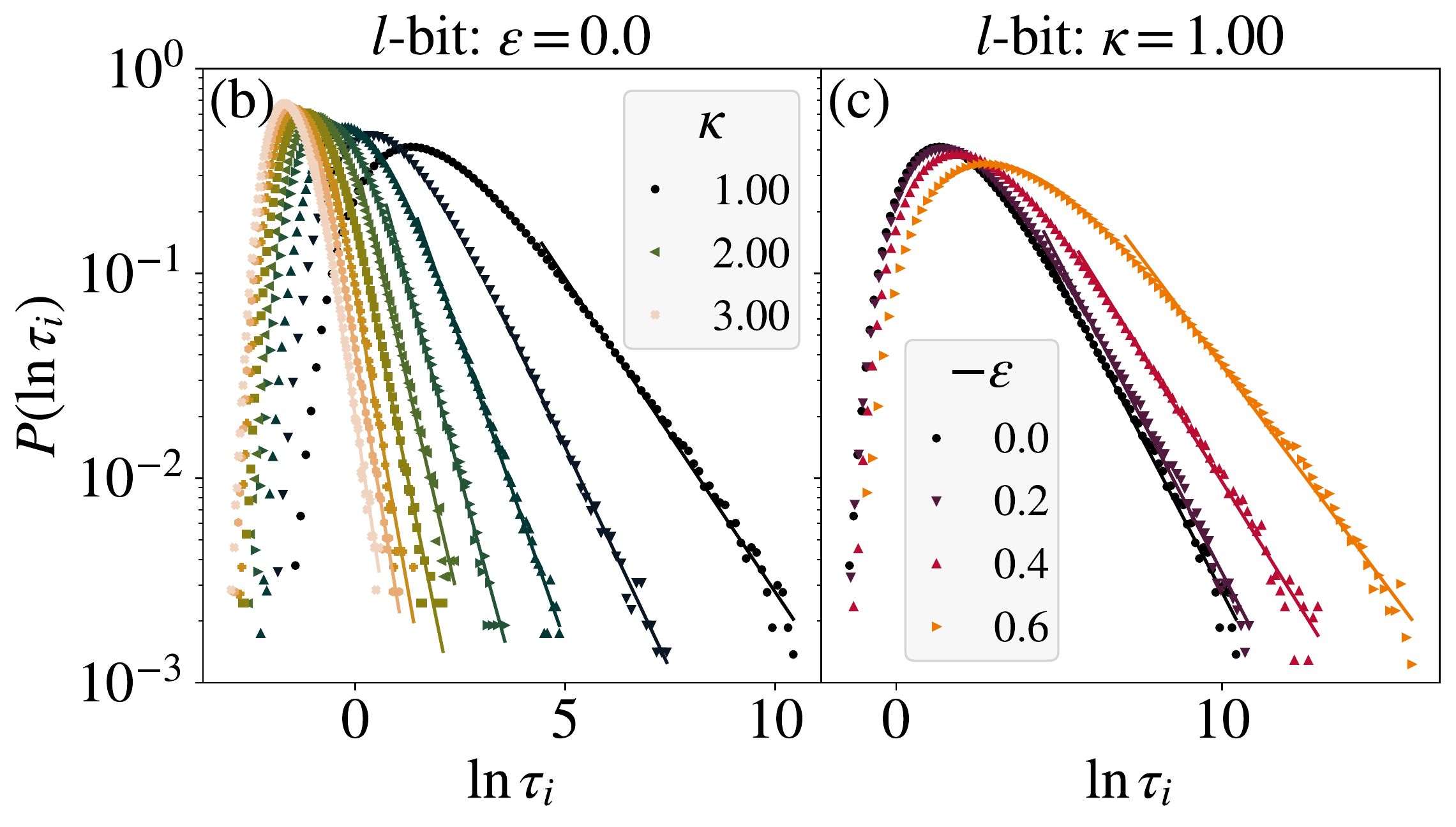}
    \label{fig:histo-tau-LBIT}}
    \caption{Probability distribution functions of $\ln(\tau_{i})$. (a)~Results for the XXZ model \eqref{eq:XXZ_model}, for $L=16$, $t_{\text{fin}}=10^3$, and various $W$. We performed the XXZ unitary dynamics using the Krylov technique~\cite{expokit}, with dimension of the  $M=40$, and at least 8000 disorder realizations. For comparison, the $l$-bit model at $L=10$, $\kappa=1$, $\varepsilon=0$ is shown as well (300 disorder realizations). $J_0$ has been fixed to make the pdf's maximum coincide with the XXZ ones. 
    (b)-(c)~Results for the $l$-bit model \eqref{eq:lbit_model} at $L=80$, for various $\kappa$ (in steps of $\Delta \kappa = 0.25$) and $\varepsilon$, averaged over at least 4000 disorder realizations and 20 initial states for each of them. As $\kappa$ decreases, i.e.\ disorder increases, the distributions broaden; the same when $\varepsilon$ decreases, in analogy with classical amorphous materials approaching the glass transition. We performed power-law fits on the tails of the pdf's, obtaining the exponents $\beta$ whose behavior is shown in Fig.~\ref{fig:powerlaw-tau}.}
    \label{fig:histo-tau}
\end{figure}

\section{Distributions of the relaxation times}
\label{sec:distributions}

We show in Fig.~\ref{fig:histo-tau} the probability distribution function (pdf) of $\ln \tau_i$, obtained within both the XXZ and the $l$-bit model. We see that within the XXZ model (Fig.~\ref{fig:histo-tau-XXZ}) the pdf's show a peak at large relaxation times, corresponding to the final simulation time of the dynamics $t_{\rm fin}$. In Appendix~\ref{sec:app_finite} we argue that this feature is due to the (typical) asymptotic value $C_i(\infty) \simeq O(2^{-L})$; see also the discussion below Eq.~\eqref{eq:def_tau}. If the time spent in such asymptotic region is too large, the relaxation time is heavily influenced by the final time of the dynamics. This is a finite-size effect, and it does disappear upon considering larger system sizes, as we show for the $l$-bit model in Appendix~\ref{sec:app_finite_lbit} (larger system sizes for the XXZ model cannot be presently considered). 

The pdf's obtained considering the $l$-bit model for $L=80$ and for different values of $\kappa$, and $\varepsilon$ are shown in Figs.~\ref{fig:histo-tau-LBIT}--\hyperref[fig:histo-tau-LBIT]{\ref*{fig:histo-tau}c}. Thanks to the large system size, these plots do not present any peak at large times, and clearly show that the the pdf of $\ln \tau_i$ has a power-law tail; thus the distribution of $\tau_i$ has a power-law tail as well. We see that the pdf's become broader as the disorder is increased (both in the XXZ and the $l$-bit model), or the energy is lowered (in the $l$-bit model).

\begin{figure}[t]
    \centering
    \subfloat[]{\includegraphics[height=0.58\columnwidth]{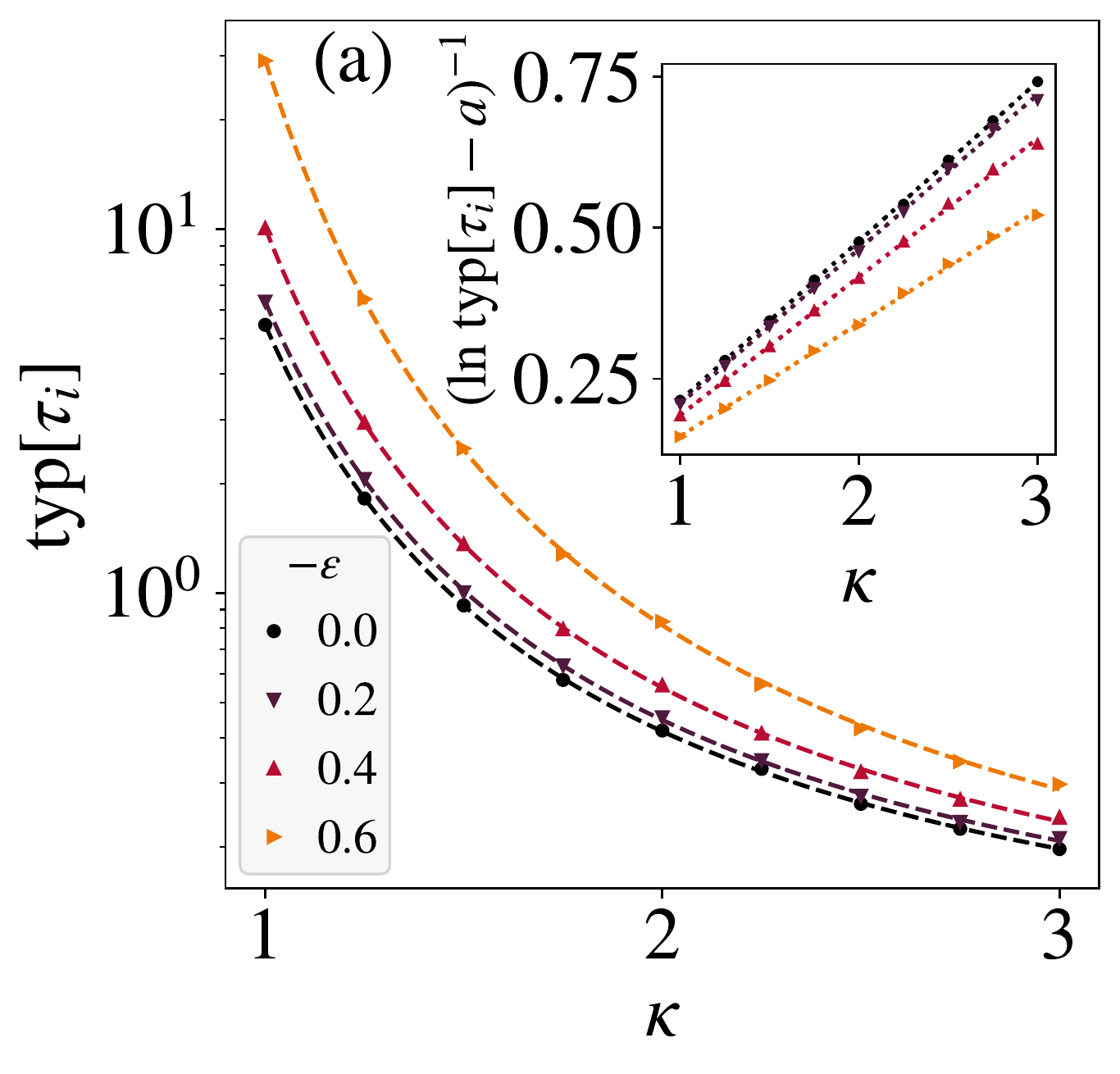}
    \label{fig:typ-tau}}
    \subfloat[]{\includegraphics[height=0.58\columnwidth]{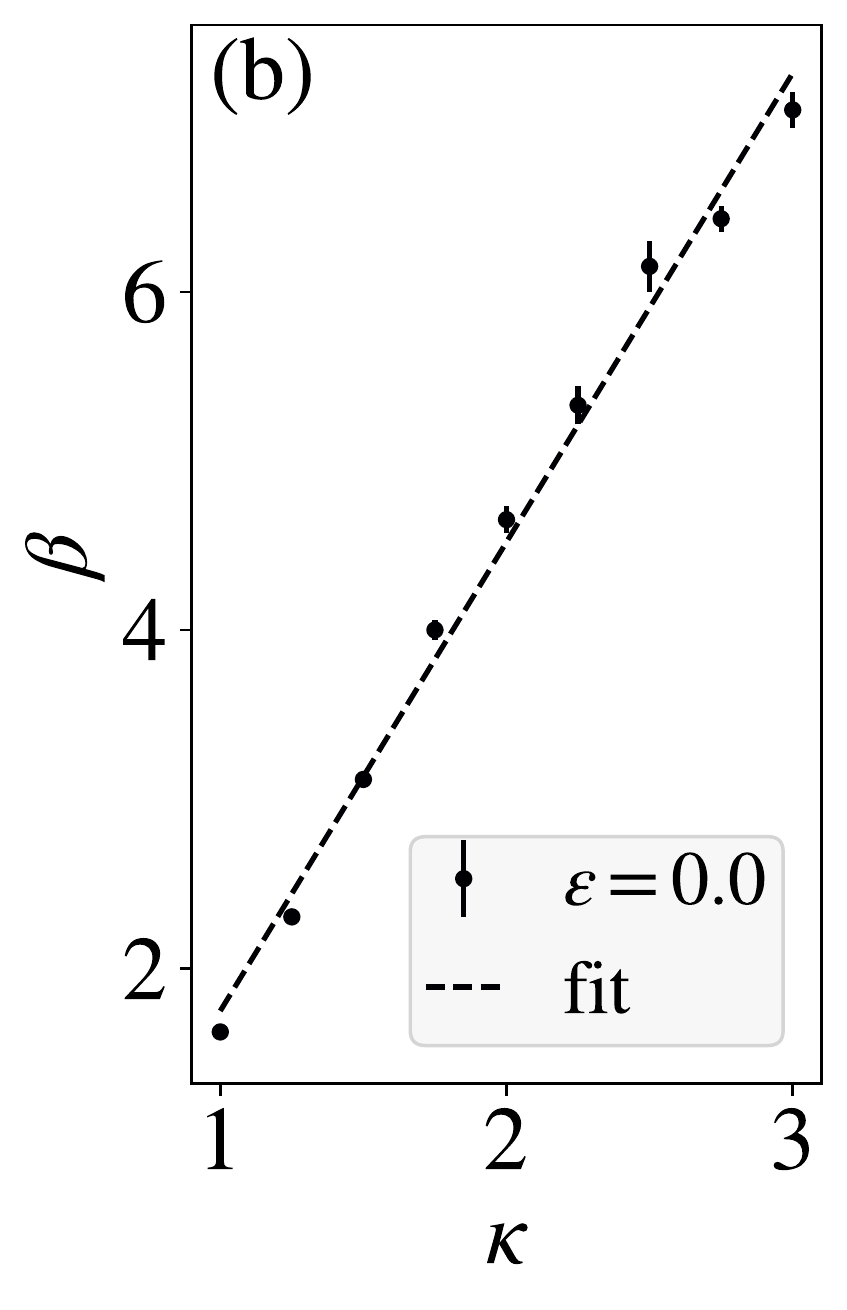}
    \label{fig:powerlaw-tau}}
    \caption{(a) The typical value of $\tau_i$, defined in the main text, is shown as a function of $\kappa$ for different $\varepsilon$ (dots). The dashed lines are fits with the function $\exp[a+(b \kappa + c)^{-1}]$. From $a$, we performed the linear fits depicted in the inset: $\ln(\typ[\tau_i]-a)^{-1}$ as a function of $\kappa$ is found to be linear, as expected (see App.~\ref{sec:app_typical_estimate}). (b)~Slope $\beta$ as a function of $\kappa$, obtained from the linear fits of the tails of $\ln P(\ln \tau_i)$ in Fig.~\ref{fig:histo-tau-LBIT}. $\beta(\kappa)$ is consistent with a linear behavior; with a linear fit we obtain: $\beta = 2.8(5) \kappa - 1.(2)$.}
\end{figure}

We define the typical value of $\tau_i$ as $\typ[\tau_i] := t_0 \exp \langle \ln(\tau_i/t_0) \rangle_{\tau_i}$, where $\langle \bullet \rangle_{\tau_i}$ is the average over the pdf of $\tau_i$. In Fig.~\ref{fig:typ-tau}, we show the behavior of $\typ[\tau_i]$ as a function of the parameters $\kappa$ and $\varepsilon$. Following usual arguments for the $l$-bit model \cite{serbyn2013universal,Swingle2013Simple,serbyn2014quantum,Znidaric18}, in Appendix~\ref{sec:app_typical_estimate} we derive the rough estimate
\begin{equation}
    \label{eq:typ_tau}
    \ln (\typ[\tau_i]/t_0) \approx (2\kappa \ln 2 - 1)^{-1}\,.
\end{equation}
Fig.~\ref{fig:typ-tau} depicts the fits of $\typ[\tau_i]$ with this functional relation with respect to $\kappa$, showing that our numerical results are in reasonable agreement with the functional form of the prediction, even if the coefficients of the fit do not match those in Eq.~\eqref{eq:typ_tau}.

In Fig.~\ref{fig:powerlaw-tau} we plot the power-law exponents $\beta$ obtained from the fit of $\ln P(\ln \tau_i) \sim -\beta(\ln\tau_i)$ shown in Fig.~\ref{fig:histo-tau-LBIT}. We see that $\beta$ has a roughly linear dependence on $\kappa$, a property which will help us interpreting the behavior of the correlation function in the next Section.

Before moving to the study of the correlation function, let us emphasize that dynamical heterogeneity is not restricted to the MBL phase. In App.~\ref{sec:app_ergodic}, we present some qualitative results also in the ergodic regime. Further investigations in this direction promise to be fruitful; however, they go beyond the scope of this work and will be discussed in forthcoming publications.

\section{Spatial correlations of the relaxation times} \label{sec:correlations}

\begin{figure}[t]
    \centering
    \subfloat[]{\includegraphics[width=\columnwidth]{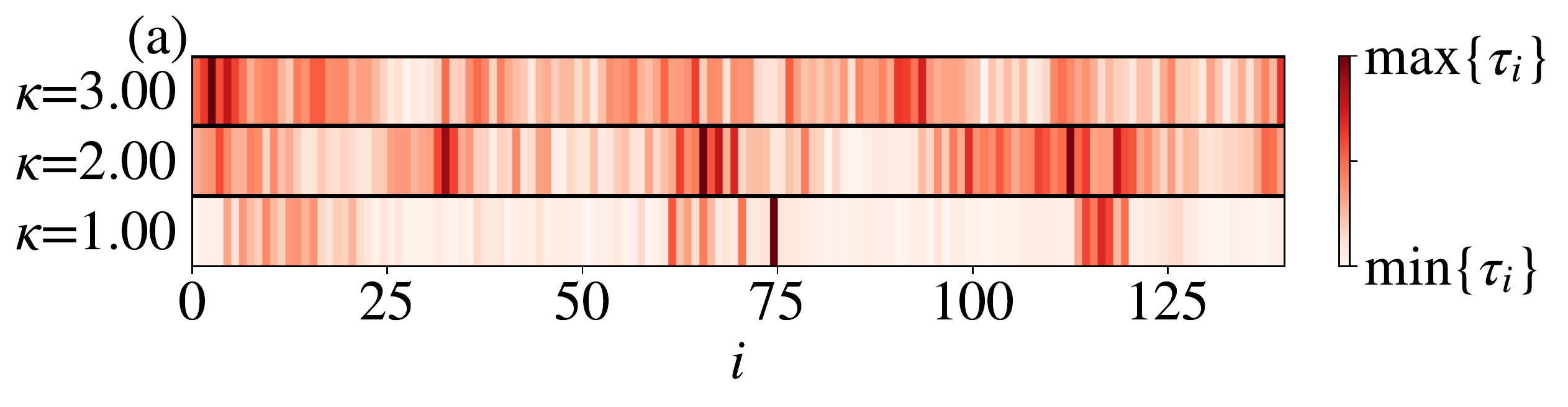}
    \label{fig:colormap-tau}}
    \vspace{-10mm}
    \subfloat[]{\includegraphics[width=\columnwidth]{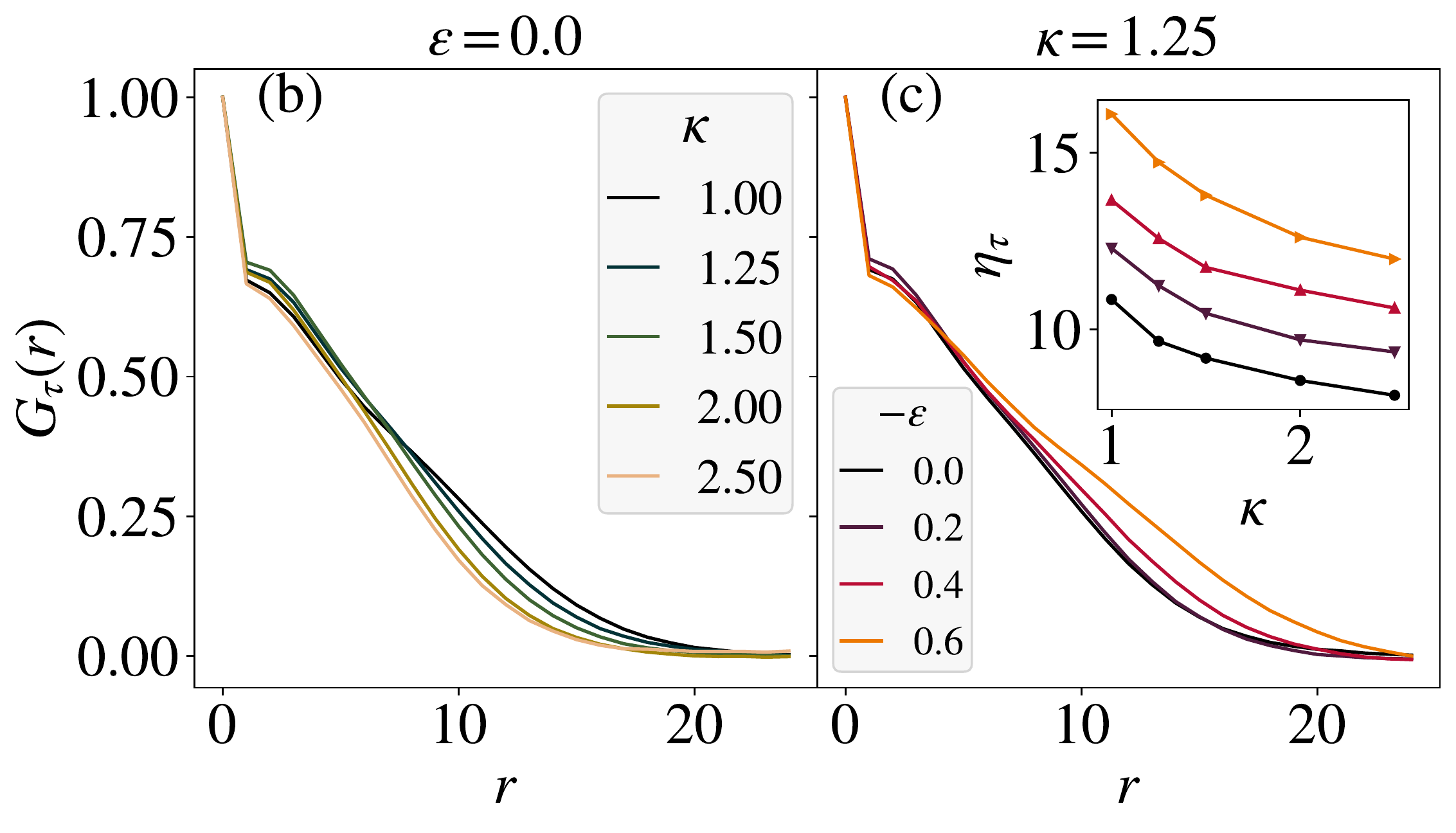}
    \label{fig:corre-tau}}
    \caption{Spatial correlation of the $\tau_i$'s in the $l$-bit model. (a)~Snapshot of the spatial distribution of the $\tau_i$'s at three different values of $\kappa$ (in one realization of disorder), showing the emergence of dynamically correlated clusters as the disorder increases. (b)--(c) The correlation function $G_{\tau}(r)$, defined in Eq.~\eqref{eq:def-Gtau}, for $L=140$, and various $\kappa$ and $\varepsilon$. We see that the spatial correlation among $\tau_i$'s increases for decreasing $\kappa$ and $\varepsilon$. Data averaged over at least 1000 disorder realizations, and 20 initial states each. (Inset of (c)) The dynamical correlation length $\eta_{\tau}$, from stretched exponential fits of $G_\tau(r)$, as a function of $\kappa$. $\eta_{\tau}$ decreases as $\kappa$ increases: for larger disorder, larger clusters of dynamically correlated spins emerge.}
    \label{fig:spatial-corre}
\end{figure}

Fig.~\ref{fig:spatial-corre} shows the spatial correlations between $\tau_i$'s. Due to the strong finite-size effects for the XXZ model, we restrict ourselves to the $l$-bit model. In Fig.~\ref{fig:colormap-tau} we report the spatial distribution of the $\tau_i$'s for a disorder realization. As $\kappa$ decreases, i.e.\ the disorder increases, the relaxation times of the local entanglement become spatially correlated over longer distances. The correlation function $G_\tau(r)$, defined in Eq.~\eqref{eq:def-Gtau}, is shown in Fig.~\ref{fig:corre-tau}--\hyperref[fig:corre-tau]{\ref*{fig:spatial-corre}c}: $G_{\tau}(r)$ decays more slowly upon decreasing $\kappa$ and $\varepsilon$, confirming the pattern observed in Fig.~\ref{fig:colormap-tau}. The same result is also supported by the (qualitative) behavior of the dynamical correlation length $\eta_{\tau}$ as a function of $\kappa$. We see in the inset of Fig.~\hyperref[fig:corre-tau]{\ref*{fig:spatial-corre}c} that $\eta_{\tau}$ decreases when $\kappa$ increases, i.e.\ when disorder decreases.

The implications are twofold. Recall that the local entanglement spreading slows down when $\kappa$ decreases (i.e.\ the disorder increases) or the energy decreases (Fig.~\ref{fig:histo-tau}). Therefore, first, increasingly larger clusters of spins emerge, in which the entanglement relaxation is correlated (Fig.~\ref{fig:spatial-corre}). Second, since the distribution of relaxation times becomes broader as disorder increases, more clusters are likely to assume an extreme value of the relaxation time in the slow, as well as in the fast tail.

These findings might seem surprising in the quantum case, as a more localized structure might be expected when disorder increases. Remarkably, a similar growth of dynamical heterogeneous clusters takes place in classical amorphous materials and spin glasses~\cite{glotzer1998dynamical}, suggesting a connection between the phenomenology close to the glass transition and to the MBL one.

We can provide an analytical argument to justify why the correlation length should increase for increasing disorder as follows. An MBL system develops increasing correlations as the time goes on, as witnessed by the entanglement entropy~\cite{montangero2006entanglement,znidaric2008many,bardarson2012unbounded,serbyn2013universal}. It is well established that, for finite systems, there is a time $t_c$ at which the system becomes fully correlated: therefore, one can roughly assume that relaxation times $\tau_i\geq t_c$ are more spatially correlated with each other than the times $\tau_i<t_c$. If, decreasing $\kappa$, the fraction of times larger than $t_c$ increases, then one can argue that the relaxation times become more correlated, in agreement with the increasing relaxation-time correlation length $\eta_\tau$ that we observe.

Let us substantiate the above argument. First, we estimate $t_c$ by quantifying the degree of correlation in the system via the half-system entanglement entropy: it increases in time as $S(t) \sim \kappa \ln(t)$~\cite{montangero2006entanglement,znidaric2008many,bardarson2012unbounded,serbyn2013universal} and, for a finite system of size $L$, it saturates to a value of order $L$ at a time $t_c \simeq \exp(L/\kappa)$. For times $\tau_i \geq t_c$, the system has reached its maximum correlation, as we have stated before. Now, let us evaluate the fraction $f_>$ of relaxation times $\tau_i > t_c$ as
\begin{equation}
    \label{eq:f>}
    f_> := \int_{\ln t_c}^\infty P(\ln \tau_i) d(\ln\tau_i),
\end{equation}
i.e.\ $f_>$ quantifies the weight of the relaxation-time distribution $P(\ln \tau_i)$ corresponding to relaxation times $\tau_i$ larger than $t_c$. Finally, we can show that $f_>$ increases with decreasing $\kappa$, meaning that the relaxation times become more correlated for increasing disorder. This simply follows from the results of Sec.~\ref{sec:distributions}, for which $\ln P(\ln \tau_i)\simeq -\beta \ln\tau_i$, with exponent $\beta = A\kappa - B$ ($A\simeq 2.8$ and $B\simeq 1$, see Fig.~\ref{fig:powerlaw-tau}). Thus, Eq.~\eqref{eq:f>} reads
\begin{equation}
  f_> = \int_{L/\kappa}^\infty \frac{1}{(\ln \tau_i)^{A\kappa - B}} d(\ln \tau_i)
  \sim \left(\frac{L}{\kappa}\right)^{1-A\kappa +B}
\end{equation}
and $f_>$ increases with decreasing $\kappa$. This means that the weight of the relaxation-time distribution associated with a maximally correlated system increases when increasing the disorder. This conclusion agrees with the behavior found for the relaxation-time correlation length, $\eta_\tau$. Finally, let us notice that the limit $\kappa\to 0$ is singular (since $t_c\to\infty$), and the argument above ceases to be valid.

\section{Conclusions}
\label{sec:conclusions}

In this work, we studied the spatio-temporal spreading of entanglement in MBL systems by monitoring the on-site concurrence. We showed that in the MBL phase the on-site concurrence behaves heterogeneously, with different functional forms and relaxation times for each site of the system. Using the tools developed for dynamical heterogeneity in classical glasses, we quantified such heterogeneous behavior of entanglement by investigating the on-site concurrence relaxation times, $\tau_i$, which display a non-trivial spatio-temporal structure.

First, we observed that the local relaxation times $\tau_i$ increase upon increasing the disorder, or upon lowering the energy of the initial states. Specifically, their distribution broadens significantly, as the exponent $\beta$, dictating the decay $P(\ln\tau_i) \sim (\ln\tau_i)^{-\beta}$ increases as $\kappa \to 0$. This can be understood in terms of the slowing down of the dynamics, due to the stronger disorder or the vicinity to the edges of the spectrum.

In addition, taking into account the spatial correlations among the $\tau_i$'s, we observed that, as disorder increases or energy decreases, increasingly larger dynamically correlated clusters arise. Within a cluster, the relaxation times are close among sites and, due to the broadness of the $\tau_i$ distribution, are likely to assume an extremely small or large value. We defined a correlation length of the relaxation times, $\eta_\tau$, which quantifies the typical extension of the correlated clusters. It represents a new length scale characterizing the MBL phase. 

The emergence of increasingly large clusters of correlated spins is somewhat surprising since one might naively expect the clusters to grow in size upon decreasing the disorder strength, i.e.\ when the delocalization transition is approached. A possible explanation lies in the fact that the concurrence quantifies only the entanglement shared by two qubits; therefore, at lower disorder, it may miss the tri-partite, or generally multi-partite, entanglement growth.

That the correlation length $\eta_\tau$ of the relaxation times increases with increasing disorder is a subtle effect due to the relaxation-time distributions having a power-law tail, with a significant weight in the long-time regime, where the system is already become fully correlated. For this point, we have provided an analytical justification.

Our analysis was mainly focused on the deep, many-body localized phase, where the $l$-bits are close to the physical spins. Similar properties for the entanglement heterogeneity are expected in all systems that present a long, localized transient before they reach a thermal state. Such systems include MBL systems coupled to a bath~\cite{nandkishore2014spectral,levi2016robustness,medvedyeva2016influence,everest2017role,nandkishore2017many,vakulchyk2018signatures,Gopalakrishnan20,wybo2020entanglement}, MBL systems in $d \geq 2$~\cite{kondov2015disorder,chandran2016many,choi2016exploring,bordia2017probing,de2017many,de2019efficiently,potirniche2019exploration,wahl2019signatures,theveniaut2020transition}, and two-level systems in structural glasses~\cite{artiaco2021signatures}. In addition, it is worth emphasizing that entanglement heterogeneity should be present also in the ergodic phase; however, we leave its characterization to future studies.

Our findings open up future research directions towards the characterization of spatio-temporal entanglement properties. A next crucial step is to explore the spatial correlation of local relaxation times in observables less affected by finite-size effects and disorder fluctuations. It would be desirable to define suitable \emph{macroscopic} observables, in the way the four-point susceptibility $\chi_4(t)$ is for classical glasses~\cite{cavagna2009supercooled,berthier2011dynamical,berthier2011dynamic,berthier2011theoretical}. In the MBL case, such observables need to detect only local entanglement fluctuations and be, possibly, experimentally measurable. 
A further interesting extension of our contribution could be to consider entanglement heterogeneity for two subsystems consisting of more than one spin, which could provide additional information on the multipartite spatio-temporal structure of quantum entanglement. Finally, we note that the observed entanglement correlated clusters, which grow in size for increasing disorder strength, might be linked to the \emph{localized bubbles} which have been the subject of numerous recent studies~\cite{de2017stability,thiery2018many}. Exploring such connections promises to be an interesting research direction for future studies.

\acknowledgements{We are grateful to A.~G.~Cavaliere, J.~H.~Bardarson, and G.~Parisi for valuable discussions. We warmly thank F.~Franzon for the careful reading of the manuscript. This work received funding from the European Research Council (ERC) under the European Union’s Horizon 2020 research and innovation program (grant agreement No.\ 101001902).}


\appendix

\section{Energy of the initial states}
\label{sec:app_initial_E}

In the $l$-bit model, given a disorder realization $\{J_{ij}\}$, we sample the local magnetization configurations $\{ m_i\} = \{\langle \sigma_i^z \rangle \}$ with probability $\propto e^{-E/T}$, $T$ being a fictitious temperature to be gradually lowered. Since $m_i \in [-1,1]$ are continuous variables, the annealing procedure has easy access to states down to the edge of the spectrum. From $\{ m_i\}$, we fix the coefficients of the initial states of the dynamics (see Eq.~\eqref{eq:lbit_initial_state}, main text) as $A_i = \sqrt{(1+m_i)/2}$, and $B_i = e^{i \phi_i}\, \sqrt{(1-m_i)/2}$. This choice guarantees that $\langle \psi_0 | H_{l\text{-bit}} | \psi_0 \rangle = E$, i.e.\ the quantum initial state has an energy expectation value equal to the desired one.

For what concerns our choice of the energy scale (see Sec.~\ref{sec:model}), namely 
\begin{equation}
    \label{eq:energy_scale}
    \varepsilon := \frac{E}{N\sqrt{h^2/3 + 2J_0^2/(e^{2/\kappa}-1)}}\,,
\end{equation}
the reasoning goes as follows. The $l$-bit Hamiltonian (Eq.~\eqref{eq:lbit_model}, main text) can be interpreted as a \emph{classical} spin glass, if one substitutes $\sigma_i^z \longrightarrow s_i = \pm 1$. Then, one can compute the (annealed) density of states of the model, finding that with high probability the ground state is at $E = - N \sqrt{h^2/4 + 4J_0^2/(e^{2/\kappa}-1)}$ (see also Ref.~\cite{derrida1981random}). Changing the spins to \emph{continuous} variables $\sigma_i^z \longrightarrow m_i \in [-1,1]$ will just modify the prefactors of $h^2$ and $J_0^2/(e^{2/\kappa}-1)$, without changing much the scale. For this reason, we have chosen to put in Eq.~\eqref{eq:energy_scale} simply the sum of the variances of $h_i$ and $\sum_j J_{ij}$. The ground state will not be exactly at $\varepsilon = -1$, but close to it.

\section{Finite-size and finite-sample effects}
\label{sec:app_finite}

\subsection{Distributions of the relaxation times in the MBL phase within the XXZ model}
\label{sec:app_finite_XXZ}

In view of the strong finite-size effects in the results for the XXZ model shown in Fig.~\ref{fig:histo-tau-XXZ}, let us better analyze the probability distribution function (pdf) of $\tau_i$.

\begin{figure}[t]
    \centering
    \subfloat[]{\includegraphics[width=78mm]{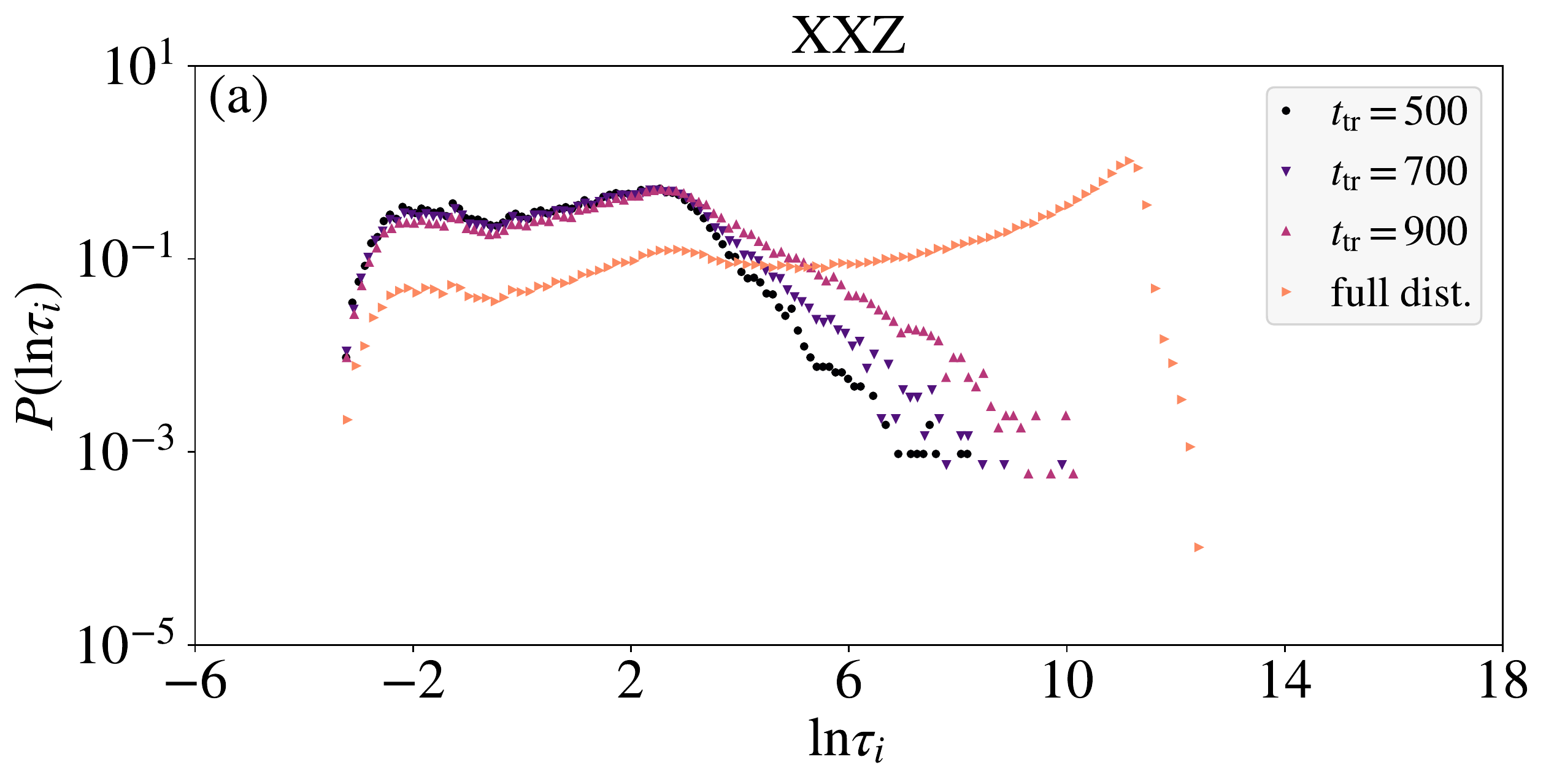}
    \label{fig:app:pdf-XXZ-W9}}
    \vspace{-9mm}
    \subfloat[]{\includegraphics[width=78mm]{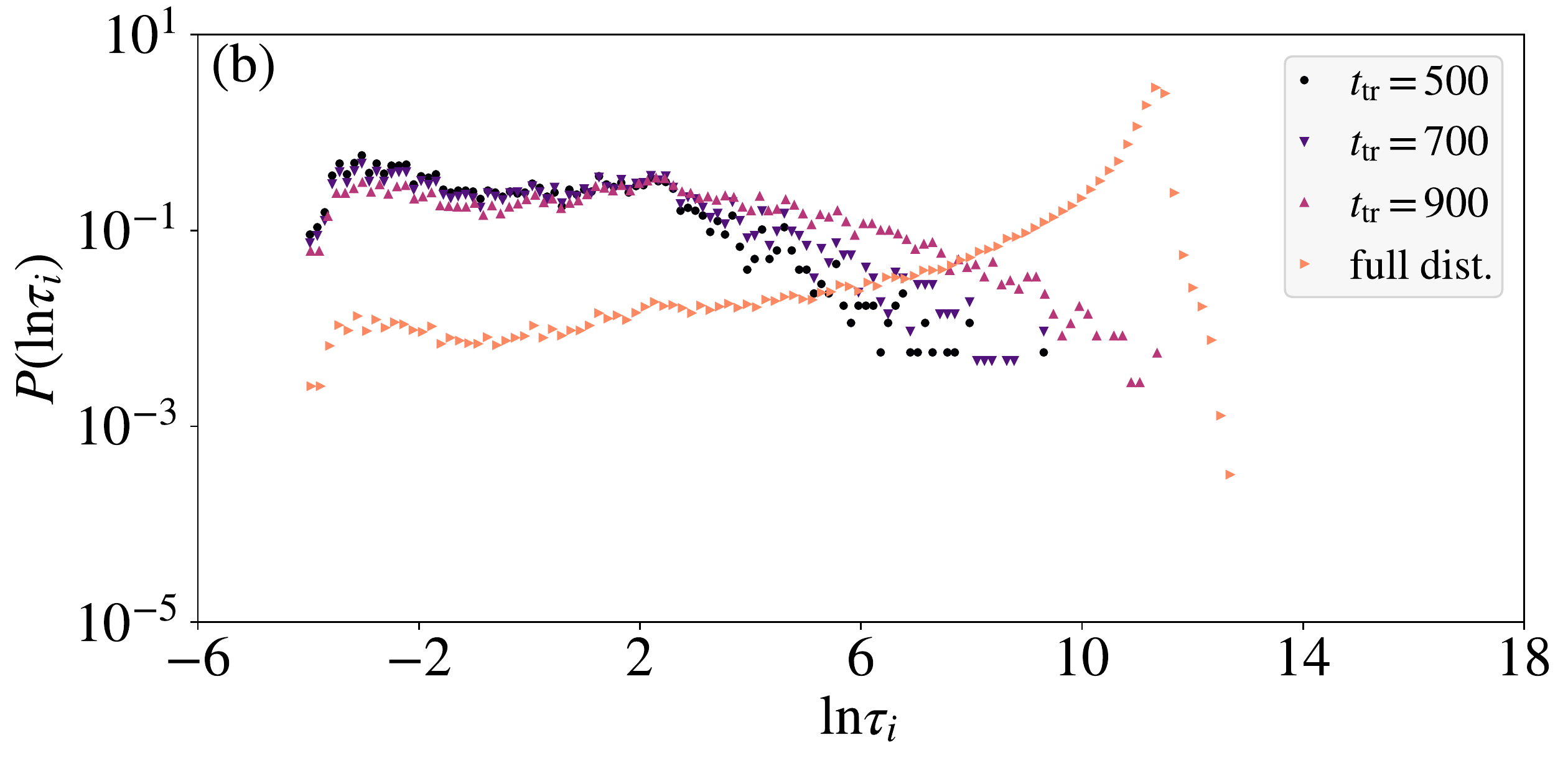}
    \label{fig:app:pdf-XXZ-W15}}
    \vspace{-9mm}
    \subfloat[]{\includegraphics[width=78mm]{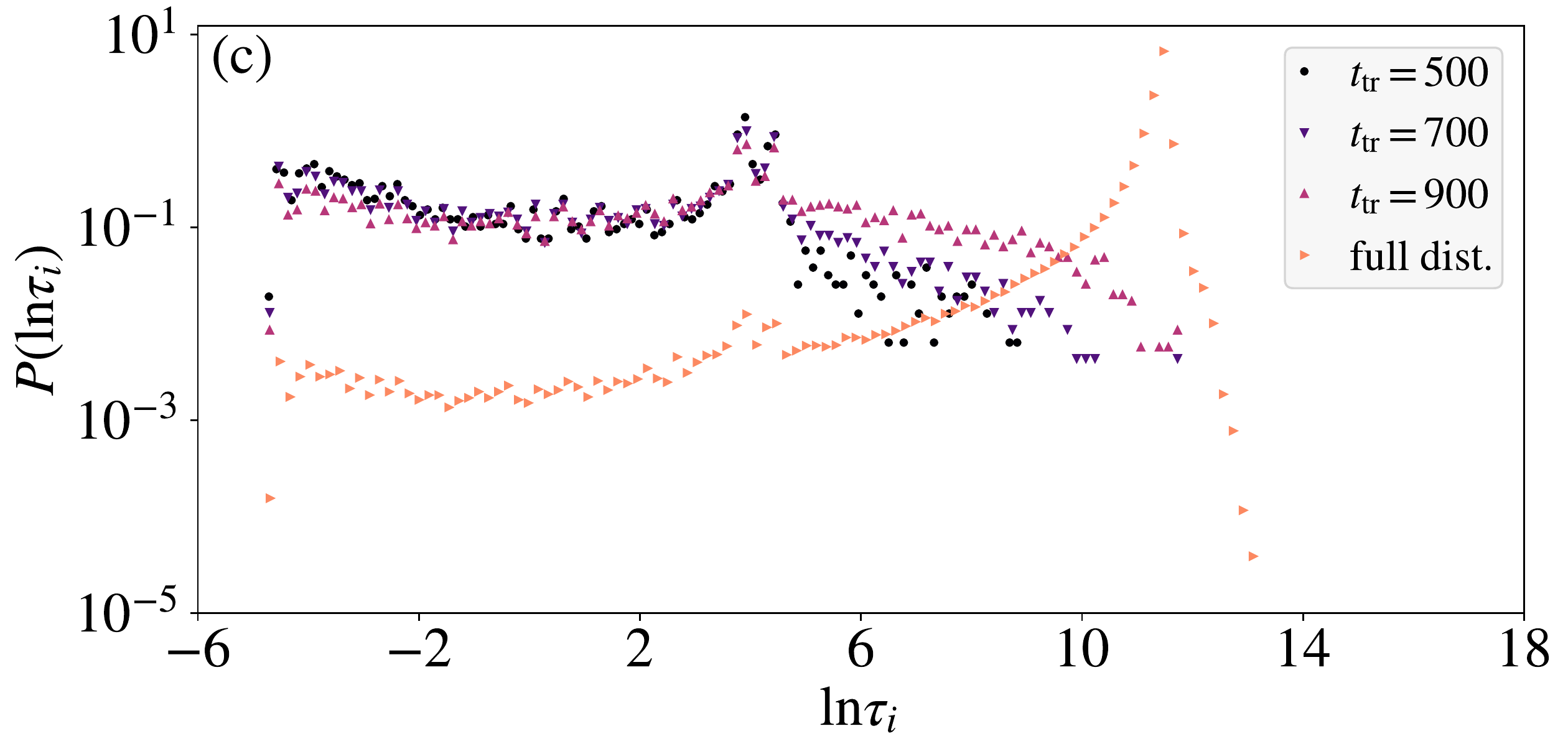}
    \label{fig:app:pdf-XXZ-W25}}
    \caption{Results for the XXZ model. We show the pdf's of  $\ln \tau_i$, truncated as described in the text. The simulations were performed with a chain of length $L=16$, final time $t_{\rm fin}=1000$, and disorder strength $W=9$ (panel (a)), $W=15$ (panel (b)), $W=25$ (panel (c)). We compare the truncated distributions with the corresponding full distribution. Data from at least 8000 disorder realizations.}
    \label{fig:app:pdf-XXZ}
\end{figure}

In Fig.~\ref{fig:app:pdf-XXZ} we show the distribution of the local relaxation times of the concurrence, computed within the XXZ model. The tail of the distribution is cut away according to the following procedure. We observe that in some instances the nearest-neighbor concurrence $C_{i,i+1}(t)$ becomes numerically indistinguishable from 0 at a time $t^*$, and then stays equal to 0 definitively. We perform an evolution lasting only a finite time $t_{\rm fin}$, so for the finite size we consider there will be many sites and realizations with $t^*>t_{\rm fin}$. This is the reason why the full distributions in Fig.~\ref{fig:app:pdf-XXZ} and Fig.~\ref{fig:histo-tau-XXZ} in the main text show such a huge peak at large times: it is formed by the contributions of $C_{i,i+1}(t)$ which have not vanished on the finite-time window $t_{\rm fin}$ of our evolution, for the finite system size we consider.

In order to get rid of this peak, we choose a certain truncation time $t_{\textrm{tr}}\leq t_{\rm fin}$, and select only the sites and the realizations for which $t^*<t_{\textrm{tr}}$. As we can see in Fig.~\ref{fig:app:pdf-XXZ} the huge peak disappears and there is a large-time tail which depends on the chosen value of $t_{\textrm{tr}}$. The small-time structure is, on the opposite, quite independent of the truncation, so we expect that it has a physical meaning. There is a peak around $\ln \tau_i \simeq -1$, which resembles the one appearing in the $l$-bit distributions; however, another peak is present around $\ln \tau_i\simeq 1$. The two-peak structure has no equal in the $l$-bit model results; we argue that this might be due to the $n$-body interactions with $n\geq3$ missing in the $l$-bit model.

\subsection{Distributions of the relaxation times within the $l$-bit model}
\label{sec:app_finite_lbit}

\begin{figure}[ht]
    \centering
    \subfloat[]{\includegraphics[width=78mm]{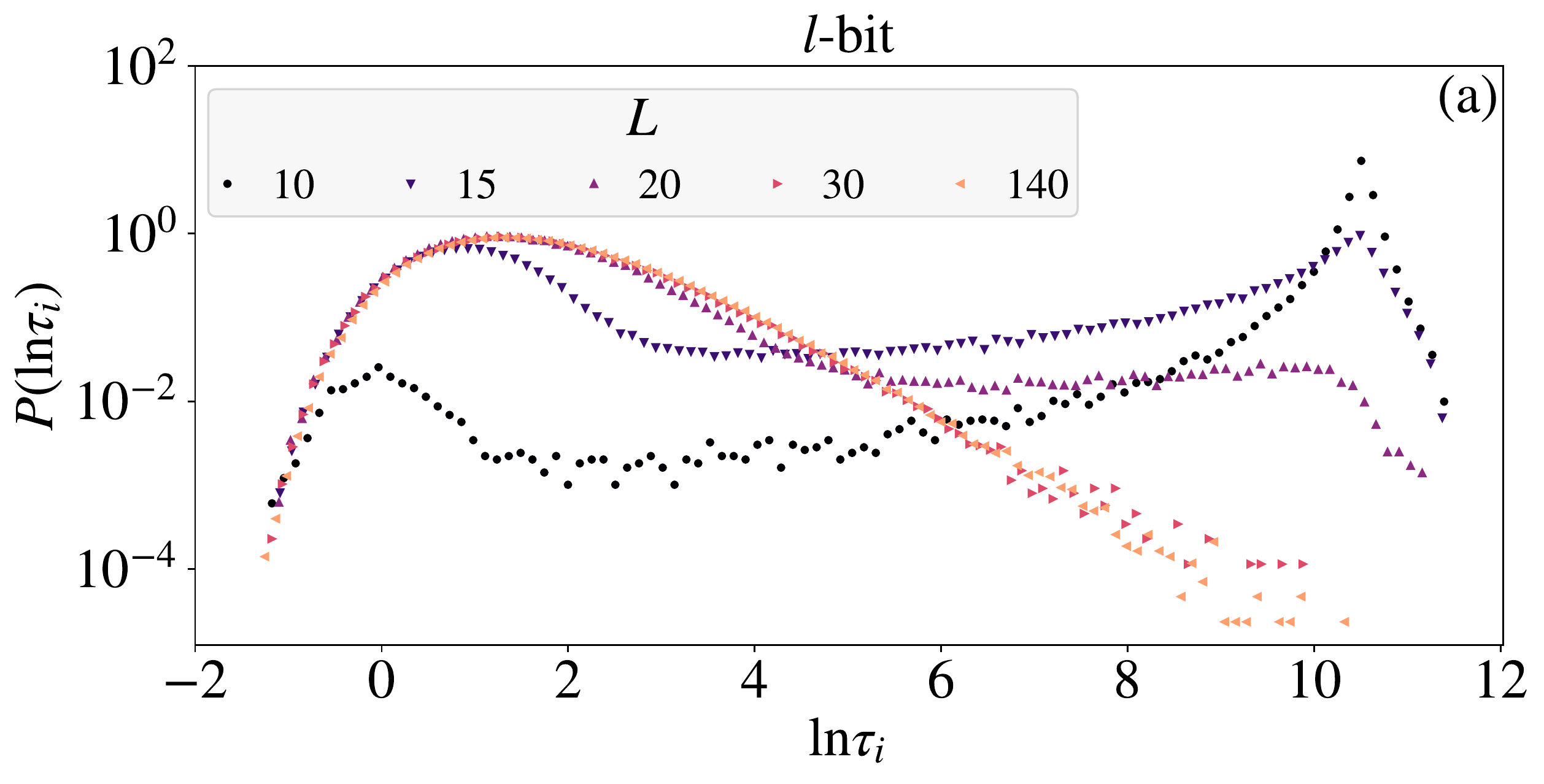}
    \label{fig:app:pdf-varyingL}}
    \vspace{-9mm}
    \subfloat[]{\includegraphics[width=78mm]{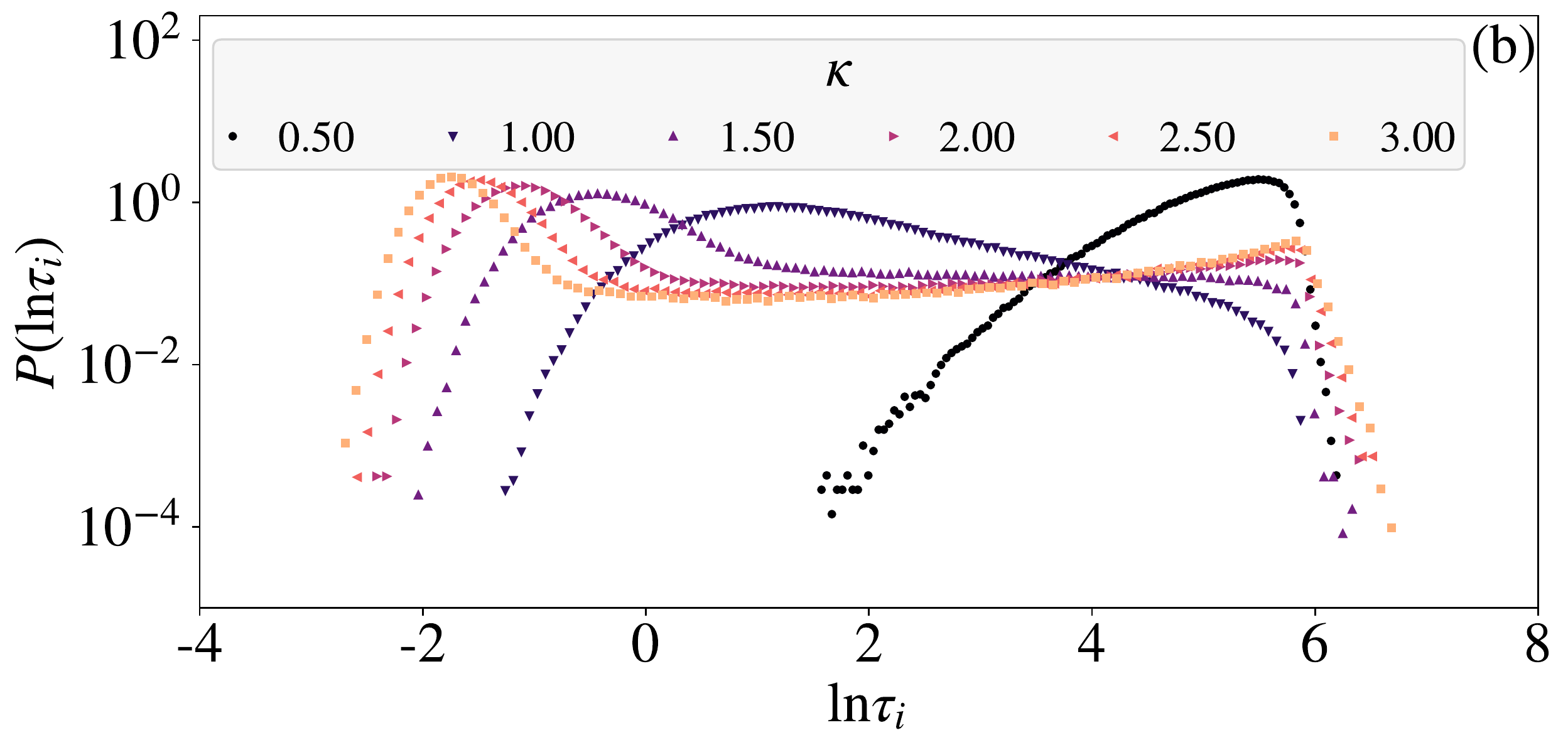}
    \label{fig:app:pdf-varyingK}}
    \vspace{-9mm}
    \subfloat[]{\includegraphics[width=78mm]{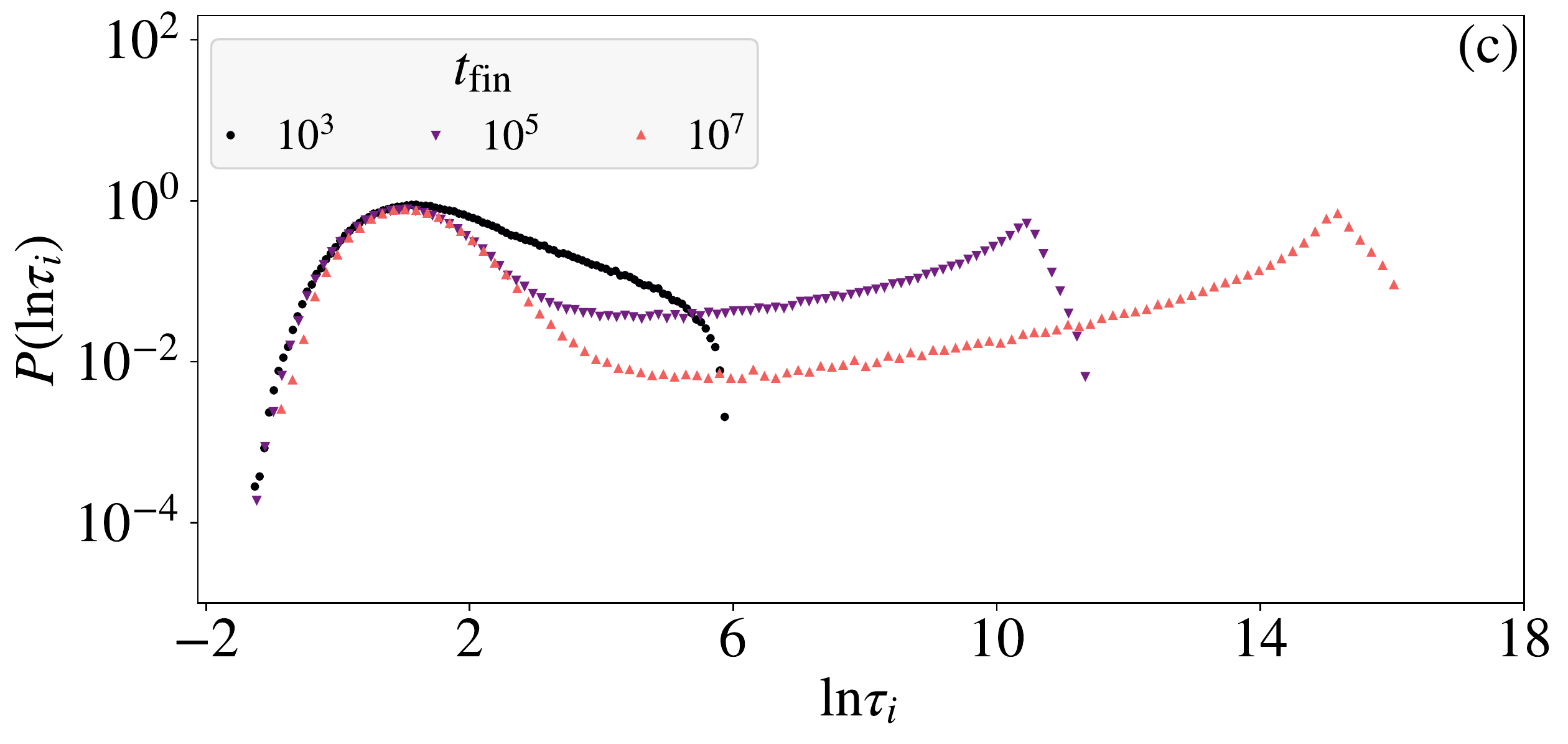}
    \label{fig:app:pdf-varyingT}}
    \caption{Pdf's of $\log_{10} {(\tau_i)}$, obtained within the $l$-bit model. (a) Results for $\kappa=1$, $\varepsilon=0$, and different values of $L$. We see that, when the system size is too small, the pdf's present a spurious peak at large values of $\tau_i$. Indeed, entanglement cannot spread over many sites, and the concurrence of some strongly interacting couples remains finite even at infinite times ($C_i(\infty) = O(2^{-L})$). We collected data from 20 initial states for, at least, 300 disorder realizations. (b) Results for $L=16$, $\varepsilon=0$, and different values of $\kappa$. Data collected from 21000 disorder realizations. (c) Results for $L=16$, $\kappa=1$, $\varepsilon=0$, and different values of the final time of the time evolution, $t_{\text{fin}}$. Data collected from 21000 disorder realizations.}
    \label{fig:app:pdf}
\end{figure}

Figure~\ref{fig:app:pdf-varyingL} shows the pdf's of $\log_{10} {(\tau_i)}$, obtained within the $l$-bit model for different values of $L$, at $\kappa=1$, $\varepsilon=0$. We see that, when $L \leq 20$, the probability distribution presents a peak at $\tau_i = O(t_{\text{fin}})$, where $t_{\text{fin}}$ is the final time used in the numerical simulations for the time evolution. This is the same effect observed in Fig.~\ref{fig:histo-tau-XXZ}. In particular, we see that the shape of the pdf's at $L=10,15,20$ strongly resembles the behavior observed in the XXZ model, confirming that those results are strongly affected by finite-size effects. 

In Figs.~\ref{fig:app:pdf-varyingK}--\ref{fig:app:pdf-varyingT} we reproduce the pdf's at $L=16$, $\varepsilon=0$, and different values of $\kappa$ an $t_{\text{fin}}$. In the presence of finite-size effects we do not observe the decay of the pdf that is found in larger system sizes. We observe instead a peak in the distribution at $\tau_i=t_{\text{fin}}$.

\section{Spatio-temporal entanglement heterogeneity in the ergodic regime}
\label{sec:app_ergodic}

\begin{figure}
    \centering
    \includegraphics[width=\columnwidth]{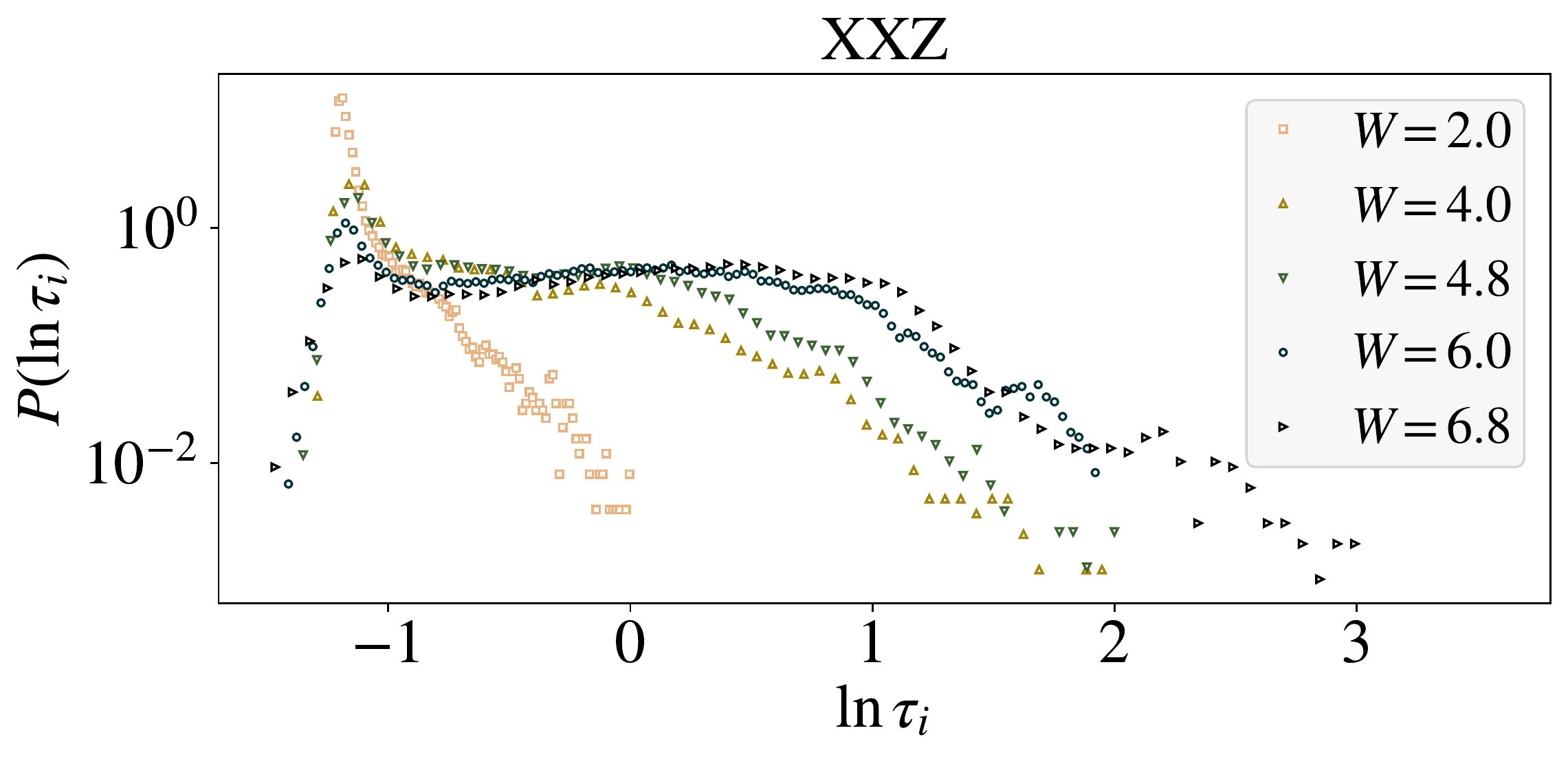}
    \caption{Distributions of the relaxation times in the XXZ model in Eq.~\eqref{eq:XXZ_model} for different values of $W$ in the ergodic phase. The peak corresponding to ergodic realizations of the on-site concurrence is very evident and decreases as the disorder increases and the transition to MBL regime is approached. Numerical parameters: $J=2$, $V=1$ (transition to MBL for $W\simeq 10$), $L=18$, at least 300 disorder realizations for each curve.}
    \label{fig:histoXXZergodic}
\end{figure}

We now present some qualitative results on entanglement heterogeneity in the ergodic regime. As shown in Fig.~\ref{fig:histoXXZergodic}, broad distributions of the relaxation times appear for the XXZ model~\eqref{eq:XXZ_model} also in the ergodic regime, and become broader as the disorder $W$ increases and the MBL-ergodicity transition is approached. Such distributions present a clear peak at small values of $\tau_i$, which decreases when disorder increases and the transition to the MBL phase is approached. This peak corresponds to realizations of the on-site concurrence that vanish very fast in time, corresponding to situations in which entanglement spreads quickly in the system. The number of such ergodic sites diminishes as the localization transition is approached, as it is quantified by the decrease in the peak height.

Notice that in the thermal phase the distributions do not show the spurious peak for large $\tau_i$, which in contrast appears in the MBL phase due to those realizations in which some $C_i(t)$ are still nonzero for $t=t_{\text{fin}}$ (contrast Fig.~\ref{fig:histoXXZergodic} with Fig.~\ref{fig:histo-tau-XXZ}).

\section{Analytical estimates of local time scales} 
\label{sec:app_typical_estimate}

Let us focus on the $l$-bit model and on the computation of typical relaxation-time scale of the ``one-site concurrence'' Eq.~\eqref{eq:one-site_concurrence}. The concurrence is a complicated non-linear function of the two-site reduced density matrix $\rho_{i,j}$, therefore it is really hard to make analytical predictions for it. However, one can hope to get a rough estimate of its behavior by considering instead the correlation function $\langle \sigma_i^x(t) \sigma_j^x(t)\rangle$. This type of correlation functions were already considered in previous works (see e.g.\ Ref.~\cite{serbyn2014quantum}), and are easy to access. Choosing $i=0$ and $j=1$ without loss of generality, it explicitly reads
\begin{widetext}
\begin{multline}
\label{eq:app_xx_correlation}
    \big \langle \sigma_0^x(t) \sigma_1^x(t) \big\rangle =
    \sum_{s_0,s_1 = \pm 1} (\rho_{0,0})_{s_0,-s_0} (\rho_{0,1})_{s_1,-s_1}
    e^{- 2i h_1 s_0 t - 2i h_2 s_1 t + 8i J_{01} s_0 s_1 t} \\
    \times \prod_{j \neq 0,1} \left[ e^{ -4i J_{0j} s_0 t -4i J_{1j} s_1 t } \cos^2 \frac{\theta_j}{2} +  e^{4i J_{0j} s_0 t +4i J_{1j} s_1 t} \sin^2 \frac{\theta_j}{2} \right],
\end{multline}
\end{widetext}
where $\rho_{0,0}$ and $\rho_{0,1}$ are the initial density matrices of sites 0 and 1, and $\theta_j$ is the azimuthal angle on the Bloch sphere for the initial state of site $j$. We take a further step, and also simplify $\theta_j \equiv \pi/2$, i.e.\ we choose a particular initial condition at infinite temperature. As a result, we find that $\langle \sigma_0^x(t) \sigma_1^x(t)\rangle$ is an oscillating function, modulated by envelopes of the form
\begin{equation}
    A_\pm(t) := \prod_{j \neq 0,1} \left| \cos \big(4 J_{0j} t \pm 4 J_{1j} t \big)\right|.
\end{equation}
It is clear that, if we want to understand the leading-order behaviour in time, we can reduce to study the simpler function
\begin{equation}
    \label{eq:prod_cos}
    A(t) := \prod_{j \neq 0} \left| \cos \big(J_{0j} t \big)\right|
\end{equation}
where, we recall, $J_{0j}$ are Gaussian variables of zero average and standard deviation $w_j := J_0 e^{-|j|/\kappa}$.

We can estimate the \emph{typical} value of $A(t)$ by means of $\typ [A(t)] := \exp \overline{\ln A(t)}$ (we need to average the \emph{logarithm} of $A$ because, with hindsight, there is a power-law tail in the relaxation-time distribution). Since
\begin{equation}\label{eq:typ}
    \overline{\ln A(t)} = \sum_j \int dJ_{0j} \, \frac{e^{-J_{0j}^2/2w^2_j}}{\sqrt{2\pi w^2_j}} \, \ln \big|\cos (J_{0j}t) \big|,
\end{equation}
we just need to compute the integral
\begin{align}
\label{eq:multi}
    \int &dJ_{0j} \, \frac{e^{-J_{0j}^2/2w^2_j}}{\sqrt{2\pi w^2_j}} \, \ln \big|\cos (J_{0j}t) \big|\nonumber \\
    &= \int dJ_{0j} \, \frac{e^{-J_{0j}^2/2w^2_j}}{\sqrt{2\pi w^2_j}} \, \left[ \ln \big| 1+ e^{2iJ_{0j}t}\big| - \ln 2 \right]\nonumber \\
    &= \int dJ_{0j} \, \frac{e^{-J_{0j}^2/2w^2_j}}{\sqrt{2\pi w^2_j}} \, \sum_{n \geq 1} \frac{(-1)^{n+1}}{n} e^{2inJ_{0j}t} - \ln 2\nonumber \\
    &= \sum_{n \geq 1} \frac{(-1)^{n+1}}{n} e^{ -2 n^2 w^2_j t^2} - \ln 2.
\end{align}

Let us first focus on the asymptotic value in time of $\typ [A(t)]$.
Substituting Eq.~\eqref{eq:multi} in Eq.~\eqref{eq:typ}, for finite system size $L$, and applying the dominated convergence theorem when performing the limit $t\to\infty$, one can show that $\typ [A(\infty)] :=\lim_{t\to\infty}\exp \overline{\ln A(t)} \simeq 2^{-L}$. This result is also related to the asymptotic value of the concurrence $\typ[ C_i(\infty)]$ being exponentially small in the system size, as discussed in Sec.~\ref{sec:methods}.

Now, we focus on finite time $t$, and we consider $L\gg 1$ so that $\typ [A(\infty)]\simeq 0$. We further proceed by approximating
\begin{equation}
    \sum_{n \geq 1} \frac{(-1)^{n+1}}{n} e^{ -2 n^2 w^2_j t^2} \approx
    \begin{cases}
        0       & w_j^2 t^2 \gtrsim 1\\
        \ln 2  & w_j^2 t^2 \lesssim 1,
    \end{cases}
\end{equation}
which implies 
\begin{equation}
    \sum_j \bigg\{ \sum_{n \geq 1} \frac{(-1)^{n+1}}{n} e^{ -2 n^2 w^2_j t^2} - \ln 2 \bigg\} \approx - N(t) \ln 2
\end{equation}
with $N(t)$ given by
\begin{equation}
    N(t) = \# \{ j \;|\; w_j^2 \, t^2 > 1 \} =
    \begin{cases}
        2 \kappa \ln \big( J_0 t \big) & t>1/J_0 \\
        0 & t<1/J_0.
    \end{cases}
\end{equation}
Finally, we find
\begin{align}
    \typ[A(t)] = \begin{cases}
        \big( J_0 t\big)^{-\kappa \ln 4} & t>1/J_0 \\
        1 & t<1/J_0.
    \end{cases}
\end{align}
Substituting this typical value in the definition of $\tau_i$ (see Eq.~\eqref{eq:def_tau}, main text), we get
\begin{align}
    \typ[\tau] &= J_0^{-1}\exp \left(\frac{\displaystyle \int_0^\infty dt \, \typ[A(t)] \, \ln(J_0 t)}{\displaystyle \int_0^\infty dt \, \typ[A(t)]}\right)\\
    &= (e J_0)^{-1}\exp \left\{ \frac{1}{\kappa \ln 4 - 1} \right\}.
\end{align}

\section{Self-averaging property of the correlation function} 
\label{self:app}

\begin{figure}[h]
    \centering
    \includegraphics[width=85mm]{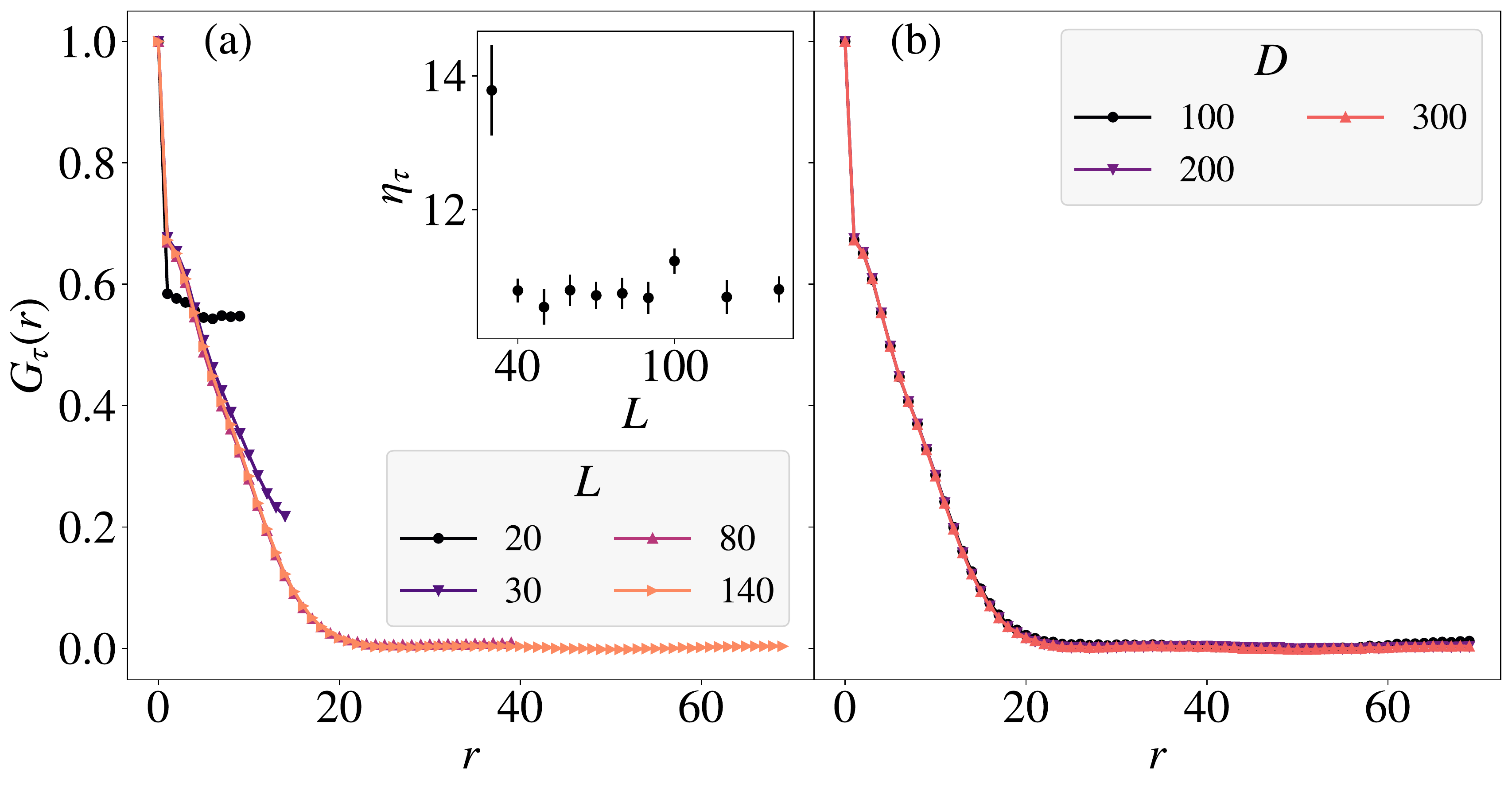}
    \caption{$G_{\tau}(r)$ defined in the main text, Eq.~\eqref{eq:def-Gtau}. Results for the $l$-bit model (see Eq.~\eqref{eq:lbit_model}, main text) for: (a) $\kappa=1$, $\varepsilon=0$, and various system sizes $L$, averaged over $D=300$ disorder realizations and 20 initial states; (b) $\kappa=1$, $\varepsilon=0$, $L=140$, averaged over a different number of disorder realizations $D$, and 20 initial states. We see that $G_{\tau}(r)$ converges quickly to its thermodynamic value (panel (a)), and is almost independent of the number of disorder realizations (panel (b)). (Inset of (a)) The dynamical correlation length $\eta_\tau$ from stretched exponential fits of $G_\tau(r)$ reported in (a) (see main text) as a function of the system size $L$. We see that $\eta_\tau$ is almost independent of the system size for $L \geq 40$. The error bars are given by the fit errors.}
    \label{fig:app:Gtau-self-averaging}
\end{figure}

As anticipated in the main text, the correlation function $G_{\tau}(r)$ (defined in Eq.~\eqref{eq:def-Gtau} of the main text) is a self-averaging quantity, as depicted in Fig.~\ref{fig:app:Gtau-self-averaging}. It is indeed only slightly sensitive to finite-size effects (see Fig.~\hyperref[fig:app:Gtau-self-averaging]{\ref*{fig:app:Gtau-self-averaging}a}), and to disorder fluctuations (see Fig.~\hyperref[fig:app:Gtau-self-averaging]{\ref*{fig:app:Gtau-self-averaging}b}). As a consequence, the dynamical correlation length $\eta_\tau$ is almost independent of the system size as well (see inset of (a)).

\bibliography{biblio}

\end{document}